\documentclass[11pt]{article}
\pdfoutput=1

\usepackage{setspace}  
\usepackage{subcaption} 
\usepackage{graphicx}
\usepackage{hyperref}
\usepackage{epsfig}
\usepackage{amsmath}
\usepackage{amsthm}
\usepackage{amssymb}
\usepackage{multirow}
\usepackage{color}
\usepackage[nosort]{cite}
\usepackage{hyperref}
\usepackage[fontsize=11.45pt]{scrextend}
\usepackage{braket,mathrsfs,slashed}
\usepackage{float}
\newlength{\abstractwidth}
\newcommand{\be}{\begin{equation}}
\newcommand{\ee}{\end{equation}}

\renewcommand{\title}[1]{\vbox{\center\bf{\Large{#1}}}\vspace{5mm}}
\renewcommand{\author}[1]{\vbox{\center#1}\vspace{5mm}}
\newcommand{\address}[1]{\vbox{\center\em#1}}

\usepackage{dsfont}

\usepackage[utf8]{inputenc}
\usepackage{rotating}
\usepackage{dsfont}
\usepackage[paper=a4paper,margin=1.5cm]{geometry}
\usepackage[utf8]{inputenc}
\usepackage{rotating}
\usepackage{soul}

\usepackage{mathrsfs} 
\usepackage{bbm}
\usepackage{etoolbox} 
\usepackage{multirow}

\renewcommand\[{\begin{equation}}
\renewcommand\]{\end{equation}}

\newcommand{\ba}{\begin{eqnarray}}
\newcommand{\ea}{\end{eqnarray}}

\hypersetup{pdfstartview=FitV,colorlinks=true,linkcolor=midblue,citecolor=midblue,filecolor=midblue,urlcolor=midblue}

\definecolor{midblue}{rgb}{0,0,0.5}

\begin{document}
	
		\newgeometry{top=3.1cm,bottom=3.1cm,right=2.4cm,left=2.4cm}
		
	\begin{titlepage}
	\begin{center}
		\vskip 0.5cm
		
		
		\title{Running EFT-hedron with null constraints \\ at loop level}

			\author{\large Guangzhuo Peng$^{a,b,c,\,\star}$, Long-Qi Shao$^{a,b,c\,\dagger}$, Anna Tokareva$^{a,d,e\,\ddagger}$, \\Yongjun Xu$^{a,b,c\,\times}$ }
			
			\address{
			$^a$School of Fundamental Physics and Mathematical Sciences, \\Hangzhou Institute for Advanced Study, UCAS, Hangzhou 310024, China\\[1.5mm]
                $^b$Institute of Theoretical Physics, Chinese Academy of Sciences, Beijing 100190, China\\[1.5mm]
                $^c$University of Chinese Academy of Sciences, Beijing 100049, China\\[1.5mm]
                $^d$International Centre for Theoretical Physics Asia-Pacific, Beijing/Hangzhou, China\\[1.5mm]
                $^e$ Department of Physics, Blackett Laboratory, Imperial College London, SW7 2AZ London, UK
                }
				\vspace{.3cm}

		\end{center}

\vspace{0.1cm}

\section*{Abstract}
Implications of general properties of quantum field theory, such as causality, unitarity, and locality include constraints on the couplings of the effective field theory (EFT) coefficients. These constraints follow from the connections between the infrared (IR) and ultraviolet (UV) theory imposed by dispersion relations for four-particle amplitudes which formally allow us to express EFT couplings through the moments of positive-definite functions (imaginary parts of partial wave amplitudes) forming the EFT-hedron geometry. Previous studies of these positivity bounds were mainly focused on the weakly coupled EFTs, limiting the analysis to tree-level amplitudes of the IR theory. In this work, we extend the scope of positivity bounds including one-loop amplitudes, which is essential especially for the loops of massless particles. Examining a single scalar theory we found that the presence of massless loops cannot be reduced only to the running of EFT couplings because loops modify the crossing symmetry relations (null constraints). Our results demonstrate that while for small coupling constants, the one-loop bounds are in good agreement with the tree-level results, the allowed EFT parameter ranges can be significantly modified if a weak coupling assumption is not additionally imposed. We present the unitarity bounds on dimension-8 and dimension-10 EFT couplings beyond the weak coupling assumption in five and six spacetime dimensions. We discuss the difficulties of obtaining the constraints in forward limit in four dimensions related to the infrared singularities and show how to overcome these problems by constructing finite combinations of null constraints.

\vspace{1cm}
\noindent\rule{6.5cm}{0.4pt}\\
$\,^\star$
\href{mailto:pengguangzhuo23@mails.ucas.ac.cn}{pengguangzhuo23@mails.ucas.ac.cn}\\
$\,^\dagger$ \href{mailto:shaolongqi22@mails.ucas.ac.cn}{shaolongqi22@mails.ucas.ac.cn}\\
$\,^\ddagger$ \href{mailto:tokareva@ucas.ac.cn}{tokareva@ucas.ac.cn}\\
$\,^\times$
\href{mailto:email2}{xuyongjun23@mails.ucas.ac.cn}\\	

\end{titlepage}

{
	\hypersetup{linkcolor=black}
	\tableofcontents
}

\baselineskip=17.63pt



\newpage

\section{Introduction}
\label{sec:intro}
Analytic properties of the scattering amplitudes can provide a possibility to connect the low energy effective field theory (EFT) with its unknown UV completion. This connection relies on causality encoded in the analytic structure of the scattering amplitudes as functions of the Mandelstam variables. Unitarity of the UV theory means the positivity of imaginary parts of partial wave amplitudes and their boundedness. Namely, in four dimensions the $2\rightarrow 2$ scattering amplitude can be expanded in eigenfunctions on angular momenta (Legendre polynomials $\mathcal{P}_j$),
\begin{equation}
    \label{pwe}
    A(s,t)=\sum_{j\,\rm{even}} n_j f_j(s)\mathcal{P}_j\left(1+\frac{2t}{s}\right)\,, \qquad n_j=16\pi (2j+1)\,,
\end{equation}

Unitarity conditions can be then formulated as $0<{\rm Im}\,f_j(s)<2$ (for massless theory).
All these properties can be summarised and used in formulating a set of dispersion relations leading to constraints on EFT Wilson coefficients.

Different approaches have been used for obtaining the constraints on EFT parameters since a seminal work \cite{Adams:2006sv}. Advances in the optimization techniques led to the conclusion that EFT coefficients are confined in compact areas after fixing the cutoff scale of the EFT \cite{Arkani-Hamed:2020blm, Bellazzini:2020cot, Tolley:2020gtv, Caron-Huot:2020cmc, Sinha:2020win, Trott:2020ebl}. The idea of constraining of the EFT parameters is based on expressing them (with the use of dispersion relations) through the sums and integrals of unknown positive-definite functions. From the point of view of mathematics, the problem of obtaining positivity bounds can then be formalized and reduced to the moment problem \cite{Bellazzini:2020cot} (see also a recent review on the corresponding mathematical results in \cite{Moment}). 

A solution to this problem produces a set of inequalities on the particular moments of the spectral density of states of the UV theory (imaginary parts of partial wave amplitudes). These inequalities imply that all values of the moments are confined in compact ranges. In other words, they lie inside a geometry called EFT-hedron in \cite{Arkani-Hamed:2020blm}. However, not all these moments can be directly related to the EFT parameters. For this reason, extra work on projecting out several (unphysical) moments is required to relate the geometry of positive moments (following \cite{Arkani-Hamed:2020blm} we will call it $a$-geometry hereafter) with bounds on EFT coefficients ($g$-geometry). Based on these foundations, systematic approaches, such as the EFT-hedron framework, have been developed recently \cite{Arkani-Hamed:2020blm,Chiang:2021ziz}, including the non-projective EFT-hedron \cite{Chiang:2022ltp} which partially incorporates full unitarity condition. Numerous applications of positivity bounds include constraints on EFTs of scalars, chiral perturbation theory, EFT of gravity, EFT of photons, Standard model EFT (SMEFT) \cite{Pham:1985cr, Pennington:1994kc,Nicolis:2009qm, Komargodski:2011vj,Remmen:2019cyz,Bellazzini:2019xts,Herrero-Valea:2019hde,Bellazzini:2017fep,deRham:2017avq,deRham:2017zjm,deRham:2017imi,Wang:2020jxr,Alberte:2020bdz, Tokuda:2020mlf,Li:2021lpe,Caron-Huot:2021rmr,Du:2021byy,Bern:2021ppb,Li:2022rag, Caron-Huot:2022ugt,Saraswat:2016eaz,Arkani-Hamed:2021ajd,Herrero-Valea:2020wxz,Guerrieri:2021ivu,Henriksson:2021ymi,EliasMiro:2022xaa,Bellazzini:2021oaj, Herrero-Valea:2022lfd,Hong:2023zgm,Chiang:2022jep,Huang:2020nqy,Noumi:2021uuv, Xu:2023lpq, Chen:2023bhu,Noumi:2022wwf,deRham:2022hpx, Hong:2024fbl,Bern:2022yes,Ma:2023vgc,DeAngelis:2023bmd,Acanfora:2023axz,Aoki:2023khq,Xu:2024iao,EliasMiro:2023fqi,McPeak:2023wmq,Riembau:2022yse,Caron-Huot:2024tsk,Caron-Huot:2024lbf,Wan:2024eto,Berman:2024owc}.

The implications of a crossing symmetry in low-energy EFT play an important role in the relation between EFT constraints and positive moments. This symmetry produces non-trivial relations between different UV integrals (moments of the spectral density of states) which are usually called null constraints \cite{Caron-Huot:2020cmc}. For the moment problem, they provide an extra set of linear relations between certain moments which, in practice, reduces the number of those moments that have to be projected out, or dimension of the EFT-hedron space. Null constraints were widely used for numerical optimization of positivity bounds \cite{Caron-Huot:2020cmc}. However, the basis of their derivation includes an assumption that the EFT amplitude can be expanded in a series of Mandelstam variables,
\begin{equation}
A(s,t)=\sum_{n=0}^{\infty}\sum_{m=0}^{\infty}g_{n,m} s^{n-m} t^m.
\end{equation}

Strictly speaking, this is valid only if the theory does not contain massless states which contribute to the amplitudes through loops. Remarkably, IR loops cannot be taken into account just as renormalization group running of the EFT Wilson coefficients \cite{Bellazzini:2021oaj}. The problems caused by the extra IR singularities sourced by loops were discussed in several recent works \cite{Bellazzini:2020cot,Bellazzini:2021oaj,Riembau:2022yse,Beadle:2024hqg,Ye:2024rzr,Bertucci:2024qzt}. 
 In \cite{Bellazzini:2021oaj,Beadle:2024hqg,Bertucci:2024qzt} dispersion relations at finite $t$ were used which allows us to avoid the problem of IR singularities. Bounds obtained numerically show that in the weakly coupled regime corrections from massless loops remain small. Attempts to extend the bounds beyond the weak coupling limit were made in \cite{Haring:2022sdp}, along the lines of the numerical bootstrap program (see, for example \cite{Paulos:2017fhb,Guerrieri:2021tak,Guerrieri:2024jkn, EliasMiro:2023fqi,EliasMiro:2022xaa,Cordova:2023wjp,Tourkine:2023xtu}
 and references therein).
 
The main purpose of this paper is to show how null constraints get modified when the contribution of loops of massless states is taken into account. Non-vanishing loop contributions to the null constraint condition were noticed in \cite{Bellazzini:2021oaj} for the case of massive particles. In our work, we show how the IR divergences can be eliminated in the massless case. We propose a  method of obtaining the compact bounds in the limit $t\rightarrow 0$ after choosing the IR-finite combinations of the arc integrals. We show that these combinations can be written in terms of positive-definite moment integrals of the density of states.

In this work, we focus on the toy model of the EFT of a single massless shift-symmetric scalar and show how the loop correction coming from its self-coupling deforms the lowest-order null constraint. We study the implications of this deformation for the EFT-hedron technique \cite{Chiang:2022ltp} of getting an upper bound on the dimension-8 and dimension-10 operator in five and six spacetime dimensions. We show that the deformation of the null constraint significantly changes the upper bound on these EFT coefficients. 

 The presence of loop corrections makes the relation between $g$-geometry and $a$-geometry non-linear, which makes $g$-geometry non-projective even if the full unitarity condition is not used. We present the deformation of the bounds on dimension-10 and dimension-12 operators and show their dependence on the Wilson coefficient of the dimension-8 operator in five and six dimensions. This effect is missing in the tree-level consideration which we found to be a valid approximation in the limit of weakly coupled EFT when dimension-8 operator is suppressed. 
 
The modification of null constraints implies that the effect of massless loops cannot be taken into account just by the proper choice of renormalization scale for the Wilson coefficients. Extra IR singularities in the forward limit in four dimensions were noticed to be problematic for the use of the same methods for constraining EFTs as for the tree-level amplitudes at $t=0$\cite{Bellazzini:2020cot,Bellazzini:2021oaj,Riembau:2022yse,Beadle:2024hqg}. From the point of view of the moment problem, IR singularities from massless loops imply that some moments correspond to divergent sums and integrals. Only certain selected combinations of them are finite and well-defined. In the weak coupling limit in four dimensions we found the two finite combinations of the first three null-constraints and show that they reproduce the same bounds as at the tree level.

The paper is organized as follows.
\begin{description}

\item[Sec.~\ref{sec:logs}:] We review the results of one-loop corrections to the scattering amplitude of the massless scalar field with shift symmetry.

\item[Sec.~\ref{sec:null}:] We review a dispersion relation technique working for the amplitudes with massless loops and find IR-finite combinations of the Wilson coefficients. We formulate modified null constraints and discuss their implications for obtaining optimal EFT constraints, as well as their interpretation in terms of geometry.
\item [Sec.~\ref{sec:g2}:] We present an upper bounds on $g_2$ and $g_3$ couplings in front of dimension-8 and dimension-10 operators in EFT derived from full unitarity conditions with the modified null constraints in five and six spacetime dimensions.
\item [Sec.~\ref{sec:g3g4}:] We obtain the constraints on higher-order couplings based on the moments technique and modified null constraint. We find that the tree-level results obtained before can get a significant corrections if loops of massless states are taken into account.
\item [Sec.~\ref{sec:conclusions}:] We comment on the implications of our results and discuss the robustness of the EFT constraints for the cases if loop corrections are non-negligible. 
\item [App.~\ref{sec:A}:] We introduce a more accurate and mathematically justified way of taking the forward limit in dispersion relations understood in the distributional sense.
\item [App.~\ref{sec:B}:] We discuss the robustness of the relation between the moment integrals, null constraints and EFT couplings, in the presence of forward limit singularities.
\item [App.~\ref{sec:C}:] We review the analytic constraints obtained as a solution to the moment problem.

\end{description}

\section{Shift-symmetric scalar field: one-loop amplitude}
\label{sec:logs}

In this paper, we show the effects of massless loops on positivity bounds using the toy model of the EFT of a single scalar field with shift symmetry. We consider four-particle scattering of identical massless real scalars. Treating all momenta as incoming, the amplitude can be written as a function of Mandelstam invariants,
\begin{equation}
s=-(p_1+p_2)^2, \quad t=-(p_2+p_3)^2, \quad u=-(p_1+p_3)^2,
\end{equation}
which satisfy $s+t+u=0$. The crossing symmetry implies that it is invariant under $s\leftrightarrow t \leftrightarrow u$ permutations
\begin{equation}
    A(s,t)=A(t,s)=A(s,u).
\end{equation}

We set the cutoff scale $\Lambda$ to unity ($\Lambda = 1$) for simplicity throughout the paper, and assume that all dimensionful quantities are expressed in units of $\Lambda$. We present our amplitude in $d$ spacetime dimensions, following the approach of~\cite{Beadle:2025cdx}. At one-loop order, the EFT amplitude can be written as follows:

\begin{equation}
\begin{aligned}
&A_{\text{low}}(s,t,u) = \ 
    g_2 \left(s^2 + t^2 + u^2\right) 
    + g_3 \, s t u 
    + g_4 \left(s^2 + t^2 + u^2\right)^2 
    + g_5 \, s t u \left(s^2 + t^2 + u^2\right) \\
    &+ \frac{t^2}{16(d^2 - 1)} \, \operatorname{Io}(t) \biggl[
        (2g_2)^2 \left( 4su - \frac{3}{2}d(3d + 2)t^2 \right)
        + 2g_2 g_3 t \left( \left( (2 - 3d)d + 8 \right)t^2 - 8su \right)
    \biggr] \\
    &+ \frac{s^2}{16(d^2 - 1)} \, \operatorname{Io}(s) \biggl[
        (2g_2)^2 \left( 4tu - \frac{3}{2}d(3d + 2)s^2 \right)
        + 2g_2 g_3 s \left( \left( (2 - 3d)d + 8 \right)s^2 - 8tu \right)
    \biggr] \\
    &+ \frac{u^2}{16(d^2 - 1)} \, \operatorname{Io}(u) \biggl[
        (2g_2)^2 \left( 4st - \frac{3}{2}d(3d + 2)u^2 \right)
        + 2g_2 g_3 u \left( \left( (2 - 3d)d + 8 \right)u^2 - 8ts \right)
    \biggr] + \dots
    \label{EFT}
\end{aligned}
\end{equation}

where 
\begin{equation}
    \operatorname{Io}(t)=\frac{i \mu^{2\epsilon} \, \Gamma(2-d/2) \, \Gamma(d/2-1)^2 }{(4\pi)^2 \, \Gamma(d-2)} \,(-\frac{t}{4\pi})^{d/2-2}.
\end{equation}

To prevent confusion in the derivation of the subsequent dispersion relation, we clarify that $\epsilon$ denotes the dimensional regularization parameter.

For future reference, we define the tree-level amplitude as:
\begin{align}
A_{\text{tree}}(s,t,u) &= 
g_2 \left(s^2 + t^2 + u^2\right) 
+ g_3 \, s t u 
+ g_4 \left(s^2 + t^2 + u^2\right)^2 \notag \\
&\quad 
+ g_5 \, s t u \left(s^2 + t^2 + u^2\right)
+ g_6 \, (s t u)^2
+ g_6' \left(s^2 + t^2 + u^2\right)^3.
\end{align}

The amplitude $A_{low}$ provides one-loop corrections to the $s^4$ and $s^5$ terms explicitly, while the running of higher-order terms is more intricate, as it involves both higher-loop contributions in the EFT and corrections from higher-point vertices. From \cite{Beadle:2025cdx}, higher-order corrections can also be obtained. However, we emphasize that the structure of IR singularities arising from light loops plays a crucial role in constraining EFT parameters through analyticity and unitarity, although we leave the analysis of more complex cases for future work.

\section{Modification of null constraints}
\label{sec:null}
The null constraints first proposed by \cite{Caron-Huot:2020cmc} reflect the properties of full crossing symmetry. In the previous works \cite{Caron-Huot:2020cmc,Chiang:2021ziz}, these constraints were derived using tree-level amplitudes and relying on the assumption that the amplitude can be expanded in series both in $s$ and $t$. In this section, we show how the null constraints get modified beyond the tree-level analysis. Namely, we found that the same combinations of integrals and sums over partial waves which were zero at tree-level become proportional to $\beta$-functions of low-energy EFT.

\subsection{Dispersion relations at loop level}
Analyticity properties of the scattering amplitudes allow us to use the dispersion relations for deriving a number of constraints connecting UV physics to IR physics. As it was proposed in \cite{Bellazzini:2020cot}, we will focus on handling the arc integrals which can be computed in the EFT and relating them to the integrals of positive-definite functions representing the UV parts.

\begin{figure}[H]  
    \centering
\includegraphics[width=0.6\textwidth]{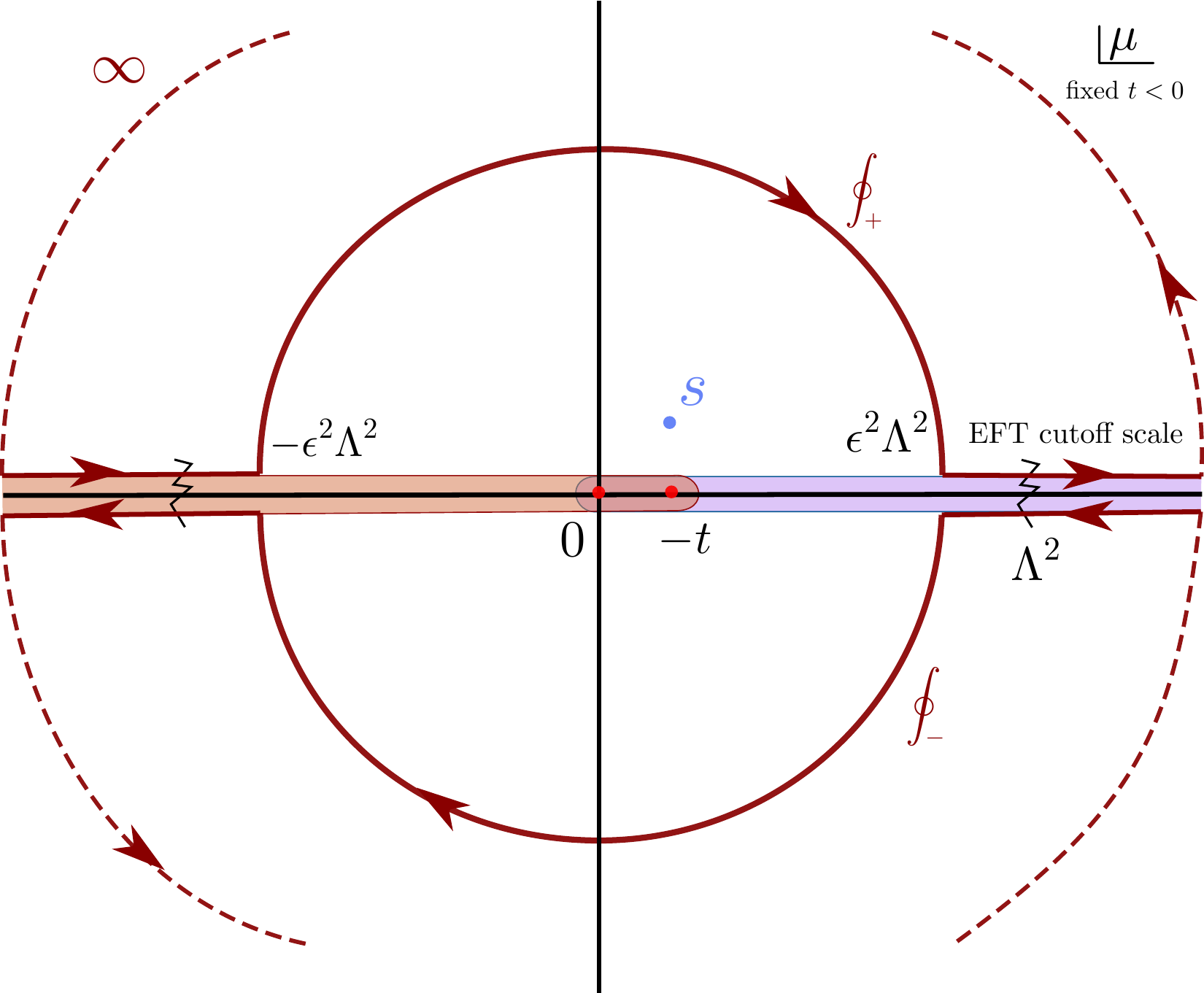}  
    \caption{ This figure shows an integration contour used in \eqref{int} and branch cut singularities of the scattering amplitude at fixed real $t<0$. Here $\Lambda$ is an EFT cutoff scale which we set to be unity, and $\epsilon<1$ parametrizes our choice of the size of the arc in \eqref{arc} which still allows to perform computations with the use of EFT.}  
    \label{fig: 1}  
\end{figure}

The Cauchy theorem applied to the contours in the upper and lower half-plane (marked + and - respectively) in Figure \ref{fig: 1} leads to:

\begin{equation}
\label{int}
    \frac{1}{2\pi i}(\oint_{+}+\oint_{-})\frac{A(\mu,t) d \mu}{(\mu-s)^{n+1}}=0,\quad n\geq 2
\end{equation}

The choice of $n+1$ as the power of the denominator will become evident in the following derivation. Consequently, the sum of the contributions from the upper and lower arcs, with the radius $\epsilon^2 \Lambda^2$  ($\epsilon < 1$, where we set units of $\Lambda=1$), is expressed as follows. The integral over the dashed contour (infinitely large arc) vanishes under the assumption that $\frac{A(s,t)}{s^2} \to 0$ as $s \to \infty$. The latter is granted by the Martin-Froissart bound \cite{Froissart:1961ux, Martin:1962rt, Jin:1964zz} in the case of the absence of massless states but it can still be treated as an implication of the locality assumption of the UV theory\footnote{See the recent discussion on violations of locality in scattering amplitudes in \cite{Tokuda:2019nqb,Buoninfante:2023dyd,Buoninfante:2024ibt}.}. Thus, if an infinite arc contribution can be neglected, the EFT arc integral can be expressed through the integrals of discontinuities of the amplitude,

\begin{equation}
\label{arc}
   \frac{1}{2\pi i} \oint_{arc}\frac{A(\mu,t)}{(\mu-s)^{n+1}}=\int_{\epsilon^2 }^{\infty} \frac{d \mu}{\pi} \frac{\operatorname{Disc}_s A(\mu, t)}{(\mu-s)^{n+1}}+ (-1)^{n} \int_{\epsilon^2 -t}^{\infty} \frac{d \mu}{\pi} \frac{\operatorname{Disc}_u A(\mu, t)}{(\mu-u)^{n+1}}.
\end{equation}

Here the discontinuities of the amplitude are defined as
\begin{equation}
\begin{split}
    &{\rm Disc}_s A(\mu,t)=\lim_{\epsilon\rightarrow0^+} \frac{1}{2i}\left[A(\mu+i\epsilon,t)-A(\mu-i\epsilon,t)\right], \\
    &{\rm Disc}_u A(\mu,t)=\lim_{\epsilon\rightarrow0^+} \frac{1}{2i}\left[A(-\mu-t+i\epsilon,t)-A(-\mu-t-i\epsilon,t)\right].
    \end{split}
\end{equation}

We used crossing symmetry for rewriting the $\mu<-t$ part of the branch cut integral through the discontinuities taken at positive $\mu$. This is an important step for further use of partial wave expansion and unitarity properties of partial wave amplitudes which hold only for $s>0$.

For convenience, we set $s=0$ and define the IR quantities $M_n(t)$ which are supposed to be computable within the EFT as

\begin{equation}
   \label{M}
   M_n(t)\equiv \frac{1}{2\pi i} \oint_{arc}\frac{A(\mu,t)}{\mu^{n+1}}+(-1)^{n} \int_{\epsilon^2 }^{\epsilon^2 -t} \frac{d \mu}{\pi} \frac{\operatorname{Disc}_s A(\mu, t)}{(\mu-u)^{n+1}}.
    \end{equation}
    
Since we can expand the functions in branch cut integrals in partial waves, we can define a set of integrals determined by the UV theory, 
    
\begin{equation}
    \label{B}
   B_n(t)\equiv \left< \frac{\mathcal{P}_j^d(1+\frac{2t}{\mu})}{\mu^{n+1}}\right>+(-1)^n\left< \frac{\mathcal{P}_j^d(1+\frac{2t}{\mu})}{(\mu+t)^{n+1}}\right>.
\end{equation}
    
Here the bracket is defined as 

\begin{equation}
\label{brackets}
   \langle X(\mu, j)\rangle\equiv \sum_{j\,\rm{even}} n_j^{(d)} \int_{\epsilon^2}^\infty  {\rm d}\mu \, \mu^{\frac{4-d}{2}} \rho_j(\mu) X (\mu, j).   
\end{equation}

where $\mathcal{P}_j^d(x)$ is the $d$ dimensional version of Legendre polynomials (Gegenbauer polynomials):

\begin{equation}
    \mathcal{P}_j^d(x) \equiv {}_2F_1(-j,j+d-3,\frac{d-2}{2},\frac{1-x}{2}).
\end{equation}

Here $n_j^{d}$ is

\begin{equation}
    n_j^{d}=\frac{(4\pi)^\frac{d}{w}(d+2j-3)\Gamma(d+j-3)}{\pi \Gamma(\frac{d-2}{2})\Gamma(j+1)}  \label{nf}
\end{equation}

The unitarity constraint is 

\begin{equation}
    0 \leq \rho_j(\mu) \leq 2  \quad \forall s>0,\forall j \: even
\end{equation}

With these definitions, the dispersion relation \eqref{arc} at $s=0$ becomes

\begin{equation}
\label{M=B}
    M_n(t)=B_n(t)
\end{equation}

In this paper, our analysis relies heavily on the amplitudes \( M_2(t) \) and \( M_4(t) \) in general \( d \)-dimensions. While the full expressions are too lengthy to present in the main text, we display \( M_4(t) \) up to order \( t^3 \) and \( M_2(t) \) up to order \( t^5 \), as we take up to four derivatives of \( M_2(t) \) and up to two derivatives of \( M_4(t) \) in our analysis. In this representation, \( \epsilon^2 \) denotes the size of the contour integration arc and should not be confused with the dimensional regularization parameter \( \epsilon \).

\begin{align}
M_2^d(t) =\; & 
2 g_2 
- t \left( g_3 + 4 t (-3 g_4 + g_5 t) \right) \notag \\
& - \frac{
    2^{-1 - 2d} g_2 \pi^{\frac{1}{2} - \frac{d}{2}} t^{d/2}
}{
    (-4 + d)(-2 + d)d\, \epsilon^4\, \Gamma\left( \frac{3 + d}{2} \right)
}
\Bigg\{
    -g_3 t^{1 - \frac{d}{2}} \epsilon^d 
    \Big[
        2 (-2 + d)d(-28 + 5d(-2 + 3d)) t^2 \notag \\
&\qquad\qquad\quad
        - 2 (-4 + d)d(-20 - 6d + 9d^2) t\, \epsilon^2 
        + (-4 + d)(-2 + d)(-4 + 3d)(2 + 3d) \epsilon^4
    \Big] \notag \\
&\qquad 
    + 4 (-4 + d)(-2 + d)d\, \pi\, (g_2 - g_3 t)\, \epsilon^4 \left( i + \cot\left( \frac{d\pi}{2} \right) \right)
\Bigg\} \notag \\
& + \frac{
    2^{-d} \pi^{-d/2}
}{
    \Gamma(3 + d)
}
\Bigg[
    - (-2 + d) d (4 + 3d) g_2 g_3\, \epsilon^{2 + d} \notag \\
&\quad
    + \frac{
        d(2 + d) g_2^2\, t^{d/2}\, \epsilon^d
    }{
        2 (-6 + d)(-4 + d)(-2 + d)\, \epsilon^6\, \Gamma\left( \frac{d}{2} \right)
    }
    \Big[
        6 (-4 + d)(-2 + d)(-4 + 5d(2 + 3d)) t^3 \notag \\
&\qquad\qquad
        - 2 (-6 + d)(-2 + d)(-4 + 9d(2 + 3d)) t^2 \epsilon^2 \notag \\
&\qquad\qquad
        + (-6 + d)(-4 + d)(-16 + 9d(2 + 3d)) t \epsilon^4 \notag \\
&\qquad\qquad
        - 6 (-6 + d)(-4 + d)(-2 + d)(2 + 3d) \epsilon^6
    \Big]
\Bigg]
\end{align}

\begin{align}
M_4^d(t) =\;&
\frac{
    2^{-2(d + 5)} (d - 2) d\, g_2\, \pi^{\frac{3}{2} - \frac{d}{2}}
}{
    3 t^2\, \Gamma\left( \frac{d + 3}{2} \right)
}
\Bigg\{
    - \frac{i \left[
        4 \left(256 + d \left( -16 + 3d (36 + d (20 + 3d)) \right) \right) g_2\, (-t)^{d/2}
        \right. } \notag \\
&\left. \quad - (d - 4)(d + 2)(-24 + d(-2 + 3d)) g_3\, t\, r^{d/2}
    \left( -\frac{r^2}{t^2} \right)^{-d/2}
    \left( \frac{r}{t} \right)^{d/2}
    \right] \notag \\
&\quad + 2 \left(256 + d \left( -16 + 3d (36 + d (20 + 3d)) \right) \right) g_2\, r^{d/2}
    \left( -\frac{r^2}{t^2} \right)^{-d/2}
    \left( -\frac{r}{t} \right)^{d/2}
    \csc\left( \frac{\pi d}{2} \right)
\Bigg\}
\end{align}

To compute \( M_n(t) \), we employ the low-energy EFT amplitude given in Eq.~(\ref{EFT}), under the assumption that the EFT remains valid up to the energy scale \( \epsilon^2 \Lambda^2 \).

We also note that taking two derivatives with respect to \( t \) of \( M_n \) leads to divergences in the forward limit \( t \to 0 \) when \( d = 4 \), due to the appearance of logarithmic terms. This poses a significant challenge in deriving the so-called \emph{null constraints}, which are crucial for bounding Wilson coefficients. While we briefly highlight this issue here, a detailed resolution will be presented later in the paper. Importantly, this problem does not occur in higher dimensions: the amplitude becomes softer as the spacetime dimension increases, and the second \( t \)-derivative remains finite. This allows us to derive meaningful bounds. For this reason, we first present results for \( d = 5 \) and \( d = 6 \), before turning to a detailed analysis of the \( d = 4 \) case.

\subsection{Null constraints in four dimensions}
If we consider only the tree-level amplitude in the form $A(s, t) \sim g_{n,m}s^{n-m} t^m$, performing the arc integral yields a term proportional to $g_{n,m}t^m$. By taking $m$ derivatives with respect to $t$ and then setting $t \to 0$, we can extract the EFT coefficient $g_{n,m}$. This way, EFT coefficients at tree level can be expressed through moments by means of taking $t$ derivatives of the brackets $B_n$ and setting $t=0$.

Since our theory is fully crossing symmetric, the coefficients \( g_{n,m} \) are not arbitrary. The amplitude must be a function of the crossing-symmetric combinations \( stu \) and \( s^2 + t^2 + u^2 \). Once we impose the twice-subtracted dispersion relation, the first nontrivial relation arises at \( n = 4 \). To determine the coefficient \( g_4 \), which corresponds to the structure \( (s^2 + t^2 + u^2)^2 \), there are three distinct ways to extract it. This is because the coefficients \( g_{4,0} \), \( g_{4,1} \), and \( g_{4,2} \), corresponding to the monomials \( s^4 \), \( s^3t \), and \( s^2t^2 \), are each proportional to \( g_4 \), with different numerical prefactors.

Specifically, we can: 

1. Set $n = 4$ in Eq.~(\ref{M}),taking no $t$-derivatives, and evaluate at $t \to 0$.  

2. Set $n = 2$ in Eq.~(\ref{M}), taking two $t$-derivatives, and evaluate at $t \to 0$.  

3. Use similar methods for other terms.  

These approaches lead to the trivial equality $g_4 = g_4$ on the left-hand side of Eq.~(\ref{M=B}).

\begin{equation}
 6 M_4(t) \mid_{t=0}  = \partial_t^2 M_2(t) \mid_{t=0}.
\end{equation}

However, the expressions within the brackets on the right-hand side of Eq.~(\ref{M=B}) differ, resulting in the null constraint as presented in \cite{Caron-Huot:2020cmc}.

\begin{equation}
\label{tree n4}
  6B_4(t) \mid_{t=0}  = \partial_t^2 B_2(t) \mid_{t=0}
  \quad  \Rightarrow \langle n_4(\mu^2,\mathcal{J}) \rangle=0, \quad n_4(\mu^2,\mathcal{J}) \equiv \frac{\mathcal{J}^2(\mathcal{J}^2-8)}{\mu^5},~ \mathcal{J}^2 \equiv j(j+1).
\end{equation}
 
However, when the loop-level amplitude is taken into account, the procedure becomes more subtle. First, taking derivatives with respect to \( t \) is complicated by the presence of logarithmic terms. More precisely, the combination \( \partial_t^2 B_2(t)\big|_{t=0} - 6 B_4(t)\big|_{t=0} \) that appears  a \( \log t \) term when  the loop-level amplitude contains, making it generally invalid to set \( t = 0 \) directly. Second, at loop level, the arc integral no longer corresponds directly to a simple derivative. Specifically, the left-hand side of Eq.~(\ref{M}) becomes \( g_4 \) plus several additional terms. Therefore, we must determine how \( g_4 \) is defined in terms of bracket structures and moment integrals. 

Here is the $n_4$ null constraint and  $g_4$

\begin{equation}
    n_4:\quad\left<\frac{{\mathcal J}^2({\mathcal J}^2-8)}{\mu^5}\right>=2 b_2 \log (\epsilon^2)-2b_2\log (-t)-3 b_2+2 c_2\epsilon^2,
\label{4dnull}
\end{equation}

\begin{equation}
    g_4=\left<\frac{1}{2\mu^5}\right>-\frac{1}{2}b_1\log \epsilon^2-\frac{1}{2}c_1 \epsilon^2 ,
\label{4dg4}
\end{equation}

where $\mathcal{J}^2\equiv j(j+1)$ as we defined above, and $b_1,b_2,c_1,c_2$ are defined by

\begin{equation}
b_1=-\frac{21g_2^2}{240\pi^2} \quad b_2=\frac{g_2^2}{240\pi^2}
\end{equation}

\begin{equation}
c_1=-\frac{g_2 g_3 }{60\pi^2} \quad c_2=-\frac{g_2 g_3}{240\pi^2}
\end{equation}

To address the infrared (IR) divergence arising from the \( \log t \) term in four dimensions, we consider subtracting it with higher-order null constraints. We start by constructing a linear combination of the \( n = 4 \) and \( n = 5 \) null constraints weighted with the corresponding $\beta$-functions to cancel the divergent \( \log t \) contribution. However, we found this combination alone is insufficient to yield a lower bound on \( g_3 \). To overcome this limitation, we introduce an additional convergent combination involving the \( n = 4 \) and \( n = 6 \) constraints. All together, these combinations allow us to derive a meaningful lower bound on \( g_3 \).

At the \( n = 6 \) level, logarithmic divergences arise not only from one-loop contributions proportional to \( g_2 g_3 \log(-t) \) and \( g_3^2 \log(-t) \), but also from two-loop terms scaling as \( g_2^3 \). Assuming the weak coupling limit \( g_2 \to 0 \), we neglect these two-loop contributions. 

From tree-level intuition, the structure of the \( n = 5 \) contribution is proportional to \( g_5\, stu(s^2 + t^2 + u^2) \). The \( n = 5 \) null constraint arises from the \( s^4 t \) and \( s^2 t^3 \) terms, as these terms share the same coefficient \( g_5 \). Therefore, by taking the first derivative with respect to \( t \) of the amplitude \( M_4 \), and combining it with the third \( t \)-derivative of \( M_2 \), we can construct the \( n = 5 \) null constraint.

The \( n = 6 \) null constraint can be obtained in a similar way. The relevant terms at this order are \( g_6 \, (stu)^2 \) and \( g_6'\, (s^2 + t^2 + u^2)^3 \). Using suitable combinations of derivatives, one can isolate contributions proportional to these structures and construct the corresponding constraint. 

When including loop-level amplitudes, we can follow a similar procedure. Specifically, by forming the combination  
\[
\frac{d^3 M_2(t)}{dt^3} - 12 \frac{d M_4(t)}{dt},
\]  
we can construct the \( n = 5 \) null constraint. However, explicit calculation shows that this expression contains two types of divergent terms: a logarithmic divergence \( \log t \) and a power-like divergence \( 1/t \). To avoid introducing the new \( 1/t \) divergence, we instead consider the following modified combination:
\begin{align}
   n_5: \quad \frac{d}{dt}[t\frac{d^3M_2(t)}{dt^3}]-12\frac{dM_4(t)}{dt}
   &=\quad \frac{d}{dt}[t\frac{d^3B_2(t)} {dt^3}]-12\frac{dB_4(t)}{dt} \notag\\
   &= \left <\frac{{\mathcal J}^2 \left[ {\mathcal J}^2 \left( 2 {\mathcal J}^2 - 43 \right) + 150 \right]}{\mu^6} \right>.
\end{align}

A similar \( \frac{1}{t} \) divergence also appears in the \( n = 6 \) case. Therefore, we define the following combination to eliminate these divergences:

\begin{align}
   n_6: \quad \frac{d^2}{dt^2}[t^2\frac{d^4M_2(t)}{dt^4}]-24\frac{d^2M_4(t)}{dt^2}
   &=\frac{d^2}{dt^2}[t^2\frac{d^4B_2(t)}{dt^4}]-24\frac{d^2B_4(t)}{dt^2} \notag\\
   &=\left < \frac{{\mathcal J}^2({\mathcal J}^2 - 12)(204 + ({\mathcal J}^2 - 32){\mathcal J}^2)}{6 \mu^7} \right>
\end{align}

Summarizing the derivation, the expressions for the null constraints at $n = 5$ and $n = 6$ are

\begin{align}
n_5 &\equiv \left < \frac{{\mathcal J}^2 \left[ {\mathcal J}^2 \left( 2 {\mathcal J}^2 - 43 \right) + 150 \right]}{\mu^6} \right> \nonumber\\
&= \frac{
-84\, g_2^2 
+ 84\, g_2 g_3\, \epsilon^2 \log(-t) 
+ 238\, g_2 g_3\, \epsilon^2 
- 84\, g_2 g_3\, \epsilon^2 \log(\epsilon^2) 
+ 2313\, g_2 g_4\, \epsilon^4 
+ 21\, g_3^2\, \epsilon^4
}{560\, \pi^2\, \epsilon^2},
\end{align}

\begin{align}
n_6 &\equiv \left< \frac{{\mathcal J}^2({\mathcal J}^2 - 12)\left[204 + ({\mathcal J}^2 - 32){\mathcal J}^2 \right]}{6 \mu^7} \right> \nonumber\\
&= \frac{1}{29400\, \pi^2\, \epsilon^4} \bigg[
-5880\, g_2^2 
- 10535\, g_3^2\, \epsilon^4 
+ 48\, g_2\, \epsilon^2 \left(245\, g_3 + 18541\, g_4\, \epsilon^2 \right) \nonumber\\
&\quad + 420\, \epsilon^4 \big(
(7\, g_3^2 - 972\, g_2 g_4)\log(\epsilon^2)
- (7\, g_3^2 + 24\, g_2 g_4)\log(-t)
+ 1992\, g_2 g_4 \log(\mu)
\big)
\bigg] .
\end{align}

With suitable choices of coefficients, we construct convergent linear combinations of the null constraints, defined as follows:

\begin{align}
\textit{n}_{45} &\equiv 2\, g_2^2\, n_5 + 36\, g_2 g_3\, n_4 \nonumber \\
&= \frac{g_2}{560\, \pi^2\, \epsilon^2} \Big[
-168\, g_2^3 
+ 224\, g_2^2 g_3\, \epsilon^2 
+ 37008\, g_2^2 g_4\, \epsilon^4 
- 1008\, g_2 g_3^2\, \epsilon^4 \nonumber \\
&\quad
- 103104\, g_2 g_3 g_4\, \epsilon^6 
+ 21\, g_3^3\, \epsilon^6
\Big], \label{n45}
\end{align}

\begin{align}
\textit{n}_{46} &\equiv 6\, n_4 \left(24\, g_2 g_4 + 7\, g_3^2 \right) - 7\, g_2^2\, n_6 \nonumber \\
&= \frac{1}{16800\, \pi^2\, \epsilon^4} \Bigg(
11760\, g_2^4 
- 23520\, g_2^3 g_3\, \epsilon^2 
+ 2 g_2^2 \Big( 
6125\, g_3^2 
- 905088\, g_2 \cdot 8 g_4 \cdot \epsilon^4 \nonumber \\
&\quad
- 840\, g_2 g_3 \Big( 
7 g_3^2 
+ 24\, g_2 \cdot 8 g_4 \cdot \epsilon^6 
+ 15 \Big( 
7 g_3^2 
- 474\, g_2 \cdot 8 g_4 \cdot \left(7 g_3^2 + 24\, g_2 \cdot 8 g_4 \right) \epsilon^8 \nonumber \\
&\quad\quad
+ 836640\, g_2^3 \cdot 8 g_4 \cdot \epsilon^4 \left( \log \epsilon - 2 \log \mu \right)
\Big) 
\Big) 
\Big) 
\Bigg) .\label{n46}
\end{align}

These combinations cancel the leading infrared divergences and yield well-defined bounds. We will now employ the method of~\cite{Chiang:2022ltp} to determine the allowed region in the \( g_3/g_2 \)–\( g_4/g_2 \) plane, with explicit analysis in dimensions \( d = 5 \) and \( d = 6 \).

With this regularization in place, we adopt the method of \cite{Chiang:2022ltp} to derive a two-dimensional allowed region in the \( g_4/g_2 \)–\( g_3/g_2 \) parameter space. The detailed analysis and numerical implementation of this procedure will be presented explicitly in the \( d = 5 \) and \( d = 6 \) cases.

A comment on the choice of $\epsilon$ is in order here. As we noticed, that the $\beta$-functions of higher order terms would also contribute to the null constraint, however, their contribution can be suppressed by the choice of somewhat smaller $\epsilon$. In a similar way, we expect that all $\beta$-functions of all EFT coefficients will contribute to the value of null constraint, and the only way to suppress them for granted is to choose small enough $\epsilon$. However, very small values of $\epsilon$ make the bounds significantly weaker, as we will show later. Choosing $\epsilon=1$ provides the strongest bounds, however, in order to trust them, one has to assume that infinite number of contributions from further $\beta$-functions and multiloop corrections indeed can be neglected for a particular EFT under consideration. This corresponds to the weak coupling assumption which was shown to hold for small values of $g_2$ \cite{Bellazzini:2017fep, Bellazzini:2016xrt}.
Thus, the bounds at the loop level become not so robust beyond weak coupling assumption, as it was also noticed before in several previous works \cite{Bellazzini:2020cot,Bellazzini:2021oaj,Haring:2022sdp,Riembau:2022yse,Beadle:2024hqg}.

In the rest of the paper, we analyze the implications of modified null constraints. In particular, we show how the modified null constraint affects the allowed region of \( g_3/g_2 \)–\( g_4/g_2 \), which is altered by choosing different values of \( g_2 \) (i.e., \( b_1, b_2 \sim g_2^2 \)) and \( \epsilon \).

\subsection{$d=5,6$ and $n=4$ null constraint}
We present here our computations for the $d=5,n=4$ null constraint and $g_2,g_3$ and $g_4$:

\begin{equation}
n_4:\quad\left<\frac{{4\mathcal J}^2({2\mathcal J}^2-21)}{15 \mu^5}\right>=\frac{g_2\sqrt{\epsilon^2}(3g_2-g_3 \epsilon^2)}{2304 \pi^2},
\label{d5null}
\end{equation}

\begin{equation}
2g_2-\frac{g_2 \epsilon^{5} (119 g_2+19 g_3\epsilon^2)}{14336 \pi ^2}= \left< \frac{2}{\mu^3} \right>,
\end{equation}

\begin{equation}
\frac{749  g_2^2 \epsilon^3} {36864 \pi^2}
+ g_3 \left(
    \frac{187 g_2 \epsilon^5}{61440 \pi^2} - 1
\right)= \left<\frac{4\mathcal{J}^2-9}{3\mu^4}\right>,
\end{equation}

\begin{equation}
    g_4=\left<\frac{2}{\mu^5}\right>+\frac{1}{4} \times\frac{g_2 \sqrt{\epsilon^2} (255 g_2+19 g_3 \epsilon^2)}{6144 \pi ^2},
    \label{d5g4}
\end{equation}

where $\mathcal{J}^2 \equiv j(j+2)$.

The $d=6,~n=4$ null constraint, $g_2,g_3$ and $g_4$ are, respectively,

\begin{equation}
n_4:\quad\left<\frac{{\mathcal J}^2({\mathcal J}^2-13)}{3 \mu^5}\right>=\frac{g_2 \epsilon^2 (2 g_2-g_3 \epsilon^2)}{13440 \pi ^3},
\label{d6null}
\end{equation}

\begin{equation}
2 g_{2} - \frac{g_{2} \, \epsilon^6 \left(60 g_{2} + 11 g_{3} \epsilon^2 \right)}{26880 \pi^3}=\left< \frac{2}{\mu^3} \right>,
\end{equation}

\begin{equation}
  \frac{19 g_{2}^2 \epsilon^4}{3840 \pi^3}
+ g_{3} \left( \frac{g_{2} \epsilon^6}{1152 \pi^3} - 1 \right) = \left< \frac{-3+\mathcal{J}^2}{\mu^4} \right>,
\end{equation}

\begin{equation}
    g_4=\left<\frac{2}{\mu^5}\right>+\frac{1}{4} \times \frac {g_2 \epsilon^2 (90 g_2 + 11 g_3 \epsilon^2)}{13440 \pi^3},
    \label{d6g4}
\end{equation}

where $\mathcal{J}^2\equiv j(j+3)$.

Higher order null constraints will contain $\beta$-functions of the next EFT couplings ($s^5,~s^6$, ...). The precise structure of IR singularities from higher loops is also important for obtaining the allowed EFT parameters with better accuracy, especially if there is no weak coupling assumption imposed. For this reason, the use of higher-order null constraints for optimizing the bounds on lower-order EFT couplings may lead to objections related to the fact that they are affected by loop corrections which may be very hard to compute. The truncated moment problem seems to provide a good optimization for the first EFT constraints where it is easy to control the loop corrections, and where higher-order null constraints are not needed. These methods work well in dimensions higher than four because the first $n=4$ null constraint is finite in the forward limit, and the higher order null constraints are not important for obtaining the bounds on the lowest order EFT couplings. This situation is different in four dimensions because all null constraints contain $\log{t}$ divergences which makes more couplings of higher order and higher loops contributing to the procedure of constraining lowest order coefficients.

\section{Upper bounds on $g_2$ and $g_3$ from full unitarity}
\label{sec:g2}
The structure of the bounds obtained in tree-level EFT is such that the ratios of all couplings, such as $g_3/g_2,~g_4/g_2$ do not depend on $g_2$ (recall that here we are dealing with dimensionless couplings after taking the cutoff scale $\Lambda=1$). In mathematical terms, this property can be expressed as the fact that $g$-geometry is projective, as well as $a$-geometry. However, if one includes full unitarity conditions for partial waves, $a$-geometry becomes non-projective \cite{Chiang:2022ltp}. In this Section, we include EFT loops and obtain the bound on $g_2$ taking into account the full unitarity of the UV theory.

 For loop-level bounds, the projective property is additionally destroyed by the loop corrections to the EFT operators, meaning that the bounds will generally depend on the value of \( g_2 \). One way to see it is to find an upper bound on \( g_2 \). Previous works \cite{Caron-Huot:2020cmc,Chiang:2022ltp} using different methods obtained the same upper bounds on \( g_2 \), relying on the full unitarity condition and the \( k = 4 \) null constraint. In our approach, we adopt the Minkowski sum method from \cite{Chiang:2022ltp}. However, in the \( d = 5,6 \) cases, the modified \( n=4 \) null constraints involve both \( g_2 \) and \( g_3 \), which necessitates performing a Minkowski sum in a three-dimensional space. This computation is analytically challenging, but the bounds on \( g_2 \) and \( g_3 \) can be obtained numerically using linear programming techniques, as discussed in \cite{Chiang:2022jep}. Therefore, in this chapter, we neglect the terms proportional to \( g_3 \) in order to derive an analytical bound on \( g_2 \).   We show that with the modified null constraint, the upper bound on \( g_2 \) is significantly changed.

We follow the approach detailed in \cite{Chiang:2022ltp}, incorporating the modification of the null constraint obtained in the section \ref{sec:null}, in contrast to the original paper where $n_4=0$ was used.

\subsection{Analytical bound on $g_2$}

 To obtain the most compact bounds, we set \( \epsilon = 1 \), corresponding to the energy scale at the cutoff, in this section. As mentioned above, we consider three specific values, \( g_3 = 0 \), \( g_3 = 4g_2 \), and \( g_3 = -10g_2 \), which partially characterize the full parameter space and allow us to avoid performing the full three-dimensional Minkowski sum. Under these approximations, we analytically evaluate a two-dimensional Minkowski sum rule involving
$a_2$ and \( n_4 \), which enables us to derive an upper bound on \( g_2 \). We will revisit this analysis later using numerical methods to obtain simultaneous bounds on both \( g_2 \) and \( g_3 \).

\quad Recall that in five dimensions, the expressions for \( a_2 \) and \( n_4 \) in terms of the moments of the spectral density are given as follows:

\begin{equation}
    a_2\equiv \left<\frac{1}{\mu^3}\right>.
\end{equation}

\begin{align*}
  a_2: \quad 
  g_2 - \frac{g_2 (119 g_2 + 19 g_3)}{2 \times 14336 \pi^2} 
  &= \left<\frac{1}{\mu^3}\right> 
  = \sum_{j=\text{even}}^{\infty} 128 (j+1)^2 \int_1^\infty \frac{\rho_j(\mu)}{\mu^{7/2}}\,d\mu \\
  &= \sum_{j=\text{even}}^{\infty} 128 (j+1)^2 \int_0^1 \rho_j(z) z^{3/2}\,dz,
\end{align*}

\vspace{-0.75em}

\begin{align}
\label{g2-n4-5d}
  \text{n}_4: \quad 
  \frac{g_2 (3g_2 - g_3)}{2304 \pi^2} 
  &= \left<\frac{4\mathcal{J}^2(2\mathcal{J}^2 - 21)}{15 \mu^5}\right> \notag \\
  &= \sum_{j=\text{even}}^{\infty} 128 (j+1)^2 \cdot \frac{4j(j+2)(2j(j+2) - 21)}{15} \int_1^\infty \frac{\rho_j(\mu)}{\mu^{11/2}}\,d\mu \notag \\
  &= \sum_{j=\text{even}}^{\infty} 128 (j+1)^2 \cdot \frac{4j(j+2)(2j(j+2) - 21)}{15} \int_0^1 \rho_j(z) z^{7/2}\,dz.
\end{align}
The expressions in six dimensions are:

\begin{equation}
    a_2\equiv \left<\frac{1}{\mu^3}\right>,
\end{equation}

\begin{align*}
  a_2: \quad 
  g_{2} - \frac{g_{2} (60 g_{2} + 11 g_{3}) }{2 \times26880 \pi^3}
  &= \left\langle \frac{1}{\mu^3} \right\rangle 
  = \sum_{\substack{j = \text{even}}}^{\infty} 
  64 \pi (j+1)(j+2)(2j+3) 
  \int_1^\infty \frac{\rho_j(\mu)}{\mu^4} \, d\mu \\
  &= \sum_{\substack{j = \text{even}}}^{\infty} 
  64 \pi (j+1)(j+2)(2j+3) 
  \int_0^1 \rho_j(z) z^2 \, dz.
\end{align*}

\vspace{-0.8em}

\begin{align}
\label{g2-n4-6d}
  n_4: ~
  \frac{2g_2^2 - g_2 g_3}{13440 \pi^3}
  &= \left\langle \frac{\mathcal{J}^2(\mathcal{J}^2 - 13)}{3\mu^5} \right\rangle \notag \\
  &= \sum_{\substack{j = \text{even}}}^{\infty} 
  64 \pi (j+1)(j+2)(2j+3) 
  \cdot \frac{j(j+3)(j(j+3) - 13)}{3}
  \int_1^\infty \frac{\rho_j(\mu)}{\mu^6} \, d\mu \notag \\
  &= \sum_{\substack{j = \text{even}}}^{\infty} 
  64 \pi (j+1)(j+2)(2j+3) 
  \cdot \frac{j(j+3)(j(j+3) - 13)}{3}
  \int_0^1 \rho_j(z) z^4 \, dz.
\end{align}

Here the full unitarity condition imposes \( 0< \rho_j < 2\), where $\rho_j$ is defined as in \eqref{brackets}. With this condition, determining the allowed region of \(a_2\) and \(n_4\) is a problem classified as the $L$-moment problem (with $L=2$), and the methods of solving it were developed about a hundred years ago \cite{Hausdorff:1923uf, akhiezer1934fouriersche, akhiezer1962some}. 

The \( L \)-moment problem can be formulated as the task of deriving constraints on a sequence of integrals, where the index \( k \) may take integer or half-integer values.
\begin{equation}
    b_k=\int_0^1 f(z) z^{k-1} dz,
\end{equation}

If the function is bounded as $0<f(z)<L$. The ratio $b_n/b_k$ ($n>k$) is known to have an extremum if $f(z)$ is chosen as follows. We select a point $0<m<1$ inside the interval of the integration and require $f(z)$ to be a step function, such as $f(z)=L$ for $0<z<m$, and $f(z)=0$ for $m<z<1$. This choice provides us with the lower bound on $b_n/b_k$. To get a upper bound, we need to consider a set of functions $f(z)=0$ for $0<z<m$, and $f(z)=L$ for $m<z<1$. Given this hint, we can get upper and lower parametric curves (parametrized by the values of $m$) in a two-dimensional plane of $(b_n,~ b_k)$.

Applying these results to finding the bounds on $(a_2,~n_4)$ plane, we can find the lower and upper boundaries for fixed values of $j$. They correspond to the following extremal configurations of \(\rho_j\):
\begin{itemize}
 \item \text{Lower boundary:}
    \[
    \rho_j(z) = 2\,\theta\left(m - z\right).
    \]
    \item \text{Upper boundary:} 
    \[
   \rho_j(z) = 2\,\theta\left(z- m\right).
    \]
\end{itemize}

Here, \( \theta(z) \) denotes the standard step function, defined as \( \theta(z) = 1 \) for \( z > 0 \) and \( \theta(z) = 0 \) for \( z < 0 \). By substituting these configurations into Eqs.~\eqref{g2-n4-5d} and \eqref{g2-n4-6d}, we obtain the parametric expressions for the lower and upper boundaries of the \( (a_2, n_4) \) region at each value of \( j \).

{\small
\begin{equation}
\label{eq:boundaries}
\begin{aligned}
  &\text{5D Lower boundary:} ~ 
  128(j+1)^2 \cdot \left(
    \frac{4}{5} m^{5/2}, \;
    \frac{4}{9} \cdot \frac{4j(j+2)(2j(j+2)-21)}{15} \cdot m^{9/2}
  \right), \\[0.5em]
  &\text{5D Upper boundary:} ~ 
  128(j+1)^2 \cdot \left(
    \frac{4}{5}(1 - m^{5/2}), \;
    \frac{4}{9} \cdot \frac{4j(j+2)(2j(j+2)-21)}{15} \cdot (1 - m^{9/2})
  \right), \\[0.8em]
  &\text{6D Lower boundary:} ~ 
  64\pi(j+1)(j+2)(2j+3) \cdot \left(
    \frac{2}{3} m^3, \;
    \frac{2}{5} \cdot \frac{j(j+3)(j(j+3)-13)}{3} \cdot m^5
  \right), \\[0.5em]
  &\text{6D Upper boundary:} ~ 
  64\pi(j+1)(j+2)(2j+3)\cdot\left(
    \frac{2}{3} (1 - m^3), \;
      \frac{2j(j+3)(j(j+3)-13)}{15} \cdot (1 - m^5)
  \right).
\end{aligned}
\end{equation}
}

In both cases, \( m \in [0,1] \).

\begin{figure}[H]
    \centering

    \begin{subfigure}
    {0.4\textwidth}
        \centering
    \includegraphics[width=\textwidth]{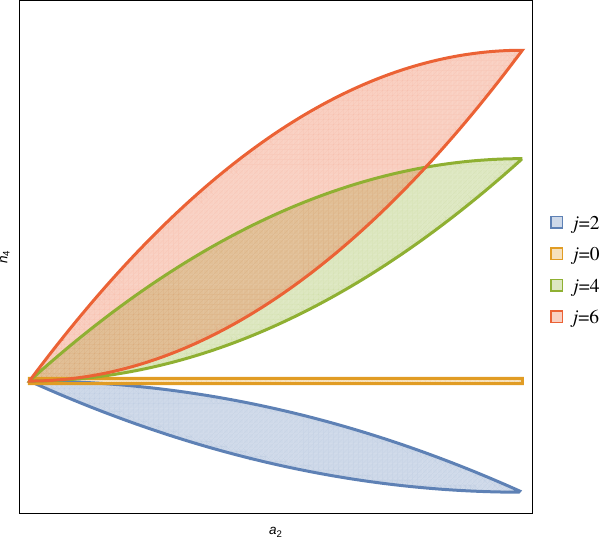}
        \caption{The rescaled regions for different values of \(j\).}
        \label{rescaled-region}
    \end{subfigure}

   \vspace{1em} 

    \begin{subfigure}{0.45\textwidth}
       \centering
        \includegraphics[width=\textwidth]{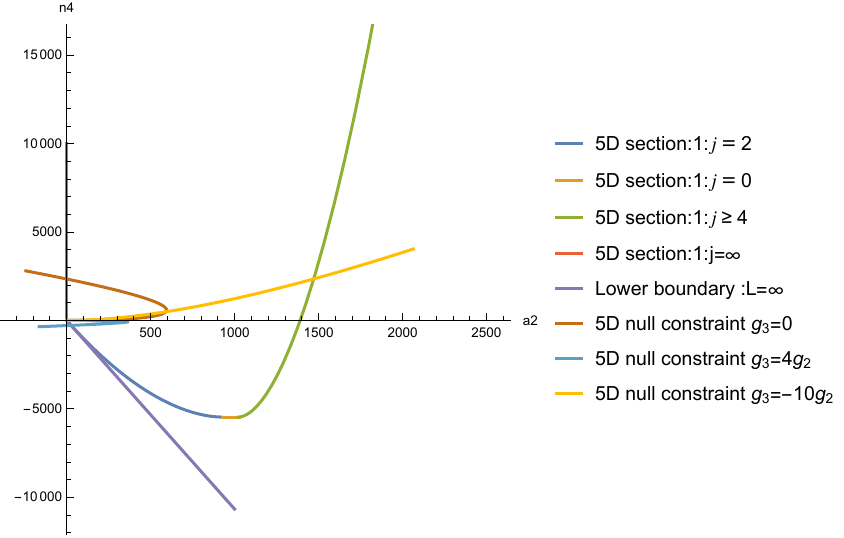}
        \caption{Allowed \((a_2, n_4)\) region in 5D.}
        \label{fig:d5-bound}
    \end{subfigure}
    \hfill
    \begin{subfigure}{0.45\textwidth}
        \centering
        \includegraphics[width=\textwidth]{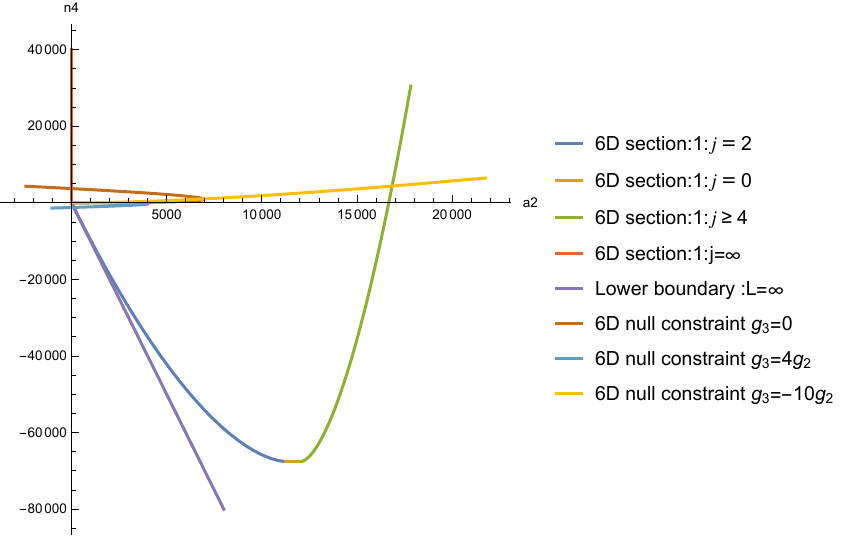}
        \caption{Allowed \((a_2, n_4)\) region in 6D.}
        \label{fig:d6-bound}
    \end{subfigure}

    \caption{
        (a) The rescaled regions for different \(j\). (b, c) The \((a_2, n_4)\)-space with four boundary sections is shown. The upper bound on \(g_2\) follows from the intersection of the lower boundary (\(j \geq 4\)) with the \(k = 4\) null constraint. The lower boundary \(L = \infty\) corresponds to the case when only positivity of partial waves is taken into account, i.e., there is no upper bound on \(\rho_j(s)\).
    }
    \label{fig:g2-n4-bounds}
\end{figure}

The shapes of these boundaries are shown in Fig. \ref{rescaled-region}. The basic idea of the elimination of $j$-dependence is based on treating the contributions from different \(j\)-values as a Minkowski sum of vectors \cite{bertsekas2003convex} in $(g_2,~n_4)$ plane. The Minkowski sum of two sets \(A\) and \(B\) in a vector space is defined as:
\[
A + B = \{ a + b \mid a \in A, \, b \in B \}.
\]

To illustrate the methods we used here, we present an example for summing up regions for $j = 0$, $j = 2$, and $j = 4$ in Fig. \ref{fig:j=0,2,4}. For better visualization, we rescaled the regions, to combine them in one plot, because their size is quickly increasing with $j$.

\begin{figure}[H]  
    \centering
    \includegraphics[width=0.4\textwidth]{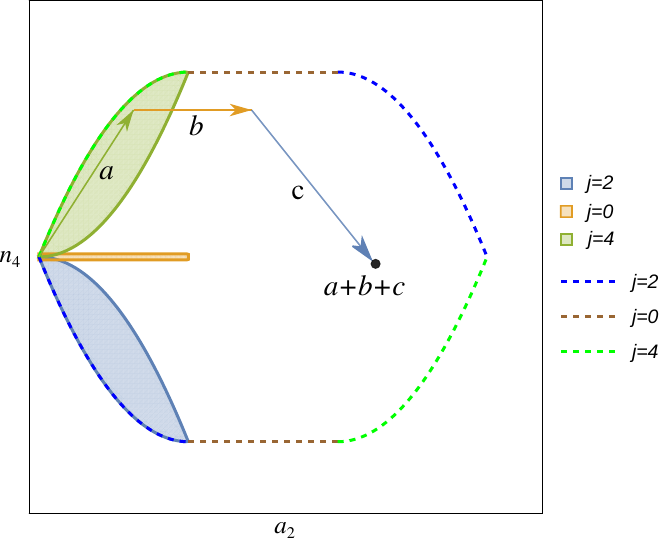}  
    \caption{ The Minkowski sum of $j=2,0,4$ regions for rescaled moments. The boundary of the sum can be obtained by ordering the individual boundaries by their slopes. }  
    \label{fig:j=0,2,4}  
\end{figure}

This method can be straightforwardly generalized to an infinite sum of regions for all values of $j$, as it is performed in detail in \cite{Chiang:2022ltp}. Instead, we directly present the final results in Fig. \ref{fig:g2-n4-bounds}.


We plot the following parametric equations corresponding to different boundary sections in \text{5D} and \text{6D}: 
\[
\begin{aligned}
\text{5D sec. (1):} ~ & a_2(m, 0, 0) = \frac{4608}{5} \left(1-m^{5/2}\right), \quad n_4(m, 0, 0) = -\frac{16384}{3}  \left(1-m^{9/2}\right); \\
\text{5D sec. (2):} ~ & a_2(0, m, 0) =\frac{512}{5} \left(m^{5/2}+9\right) , \quad n_4(0, m, 0) =-\frac{16384}{3} ; \\
\text{5D sec. (3):} ~ & a_2(0, 1, m) =1024+2560Fm^{5/2} , ~ n_4(0, 1, m) = -\frac{16384}{3}+245760Fm^{9/2},
\end{aligned}
\]
where \( F = 1.47 \). 

And the boundaries in 6D are:
\[
\begin{aligned}
\text{6D sec. (1):} ~ & a_2(m, 0, 0) = 3584 \pi  \left(1-m^3\right), \quad n_4(m, 0, 0) = -21504 \pi  \left(1-m^5\right); \\
\text{6D sec. (2):} ~ & a_2(0, m, 0) =3584 \pi +256 \pi  m^3 , \quad n_4(0, m, 0) =-21504 \pi ; \\
\text{6D sec. (3):} ~ & a_2(0, 1, m) =3840 \pi +14080 \pi F m^3 , \quad n_4(0, 1, m) = -21504 \pi +1182720 \pi F m^5,
\end{aligned}
\]

where $F=1.43$. 

From Fig.~\ref{fig:g2-n4-bounds}, we observe that the behavior of the null constraints differs significantly for the three specific values of \( g_3 \) considered. Recall the definition \( a_2 \equiv \left\langle \frac{1}{\mu^3} \right\rangle \), which depends on both \( g_2 \) and \( g_3 \). When we choose particular values such as \( g_3 = 0 \) and \( g_3 = 4g_2 \), the quantity \( a_2 \) becomes solely a function of \( g_2 \). In these cases, the relation between \( n_4 \) and \( a_2 \) takes the form of a curve that opens toward the negative \( a_2 \)-axis. We find that the null constraints do not reach the boundary of the \( (a_2, n_4) \)-space. As a result, the only effective constraint on \( g_2 \) arises from the positivity condition on \( a_2 \):
\[
\label{a2 5d}
2g_2 - \frac{g_2 \, \epsilon^{5} (119 g_2 + 19 g_3 \epsilon^2)}{14336 \pi^2} = \left\langle \frac{2}{\mu^3} \right\rangle \geq 0 \quad \text{(5D)},
\]
\[
\label{a2 6d}
2g_2 - \frac{g_2 \, \epsilon^{6} (60 g_2 + 11 g_3 \epsilon^2)}{26880 \pi^3} = \left\langle \frac{2}{\mu^3} \right\rangle \geq 0 \quad \text{(6D)},
\]
in the cases \( g_3 = 0 \) and \( g_3 = 4g_2 \).

However, when \( g_3 = -10g_2 \), the null constraints do reach the boundary of the \((a_2, n_4)\)-space. In this case, the maximum allowed value of \( g_2 \) can be determined from the intersection point of the boundary curve. Consequently, the allowed region for \( g_2 \) becomes more compact compared to the case where we rely solely on Eq.~\ref{a2 5d} and Eq.~\ref{a2 6d}.

\begin{table}[htbp]
    \centering
    \begin{tabular}{|c|c|c|}
        \hline
        Max $g_2$  & 5D & 6D \\
        \hline
        $g_3=0$ & $2375.58 $& $27739.4$\\
        \hline
        $g_3=4g_2$ & $1451.1$ & $16206.4$ \\
        \hline
        $g_3=-10g_2$ & $1471.11$ & $16769.8$ \\
        \hline
    \end{tabular}
    \caption{The maximun values of $g_2$ in different cases.}
    \label{tab:The maximun values of $g_2$ in different cases}
\end{table}

We summarize the maximum allowed values of \( g_2 \) for different choices of \( g_3 \) across various spacetime dimensions in Table~\ref{tab:The maximun values of $g_2$ in different cases}. Exceeding these values indicates a violation of the full unitarity condition. Interestingly, the positivity condition on partial waves is violated only at even larger values of \( g_2 \). In Fig.~\ref{fig:g2-n4-bounds}, this threshold corresponds to the intersection between the null constraint curve and the lower boundary associated with \( j \geq 4 \). Alternatively, it can be determined directly from the positivity condition on the bracket \( \left\langle \cdots \right\rangle \).

At tree level, there is no upper bound on \( g_2 \) arising solely from the positivity of partial waves (\( \rho_j > 0 \)), as the \( L = \infty \) line does not intersect the \( n_4 = 0 \) axis. When full unitarity is imposed, the bounds on \( g_2 \) in four dimensions were established in \cite{Chiang:2022ltp}. We found that incorporating loop-level corrections enlarges the allowed region for \( g_2 \), a feature that becomes evident in the next section. This highlights the essential role of loop effects and null constraints in deriving meaningful bounds on EFT Wilson coefficients.

This example demonstrates the significance of loop corrections for enforcing unitarity in effective field theory. Our findings suggest that, in the strong coupling regime, neglecting loop effects leads to qualitatively different unitarity constraints, underscoring the necessity of including quantum corrections for a complete picture.

\subsection{Numerical bounds on $g_2$ and $g_3$}
The numerical approach to solving the \( L \)-moment problem is quite simple and straightforward. We begin by discretizing the \( z \)-variable appearing in the definition of the moments \( a_k \). Specifically, we approximate the integral by a finite sum:
\begin{equation}
    a_k = \int_0^1 f(z)\, z^{k-1} \, dz \approx \sum_{i=1}^{N} f_{\left(\frac{i}{N}\right)}
 \left(\frac{i}{N}\right)^{k-1} \frac{1}{N}, \qquad 0 < f_{\left(\frac{i}{N}\right)} < 2.
\end{equation}

We now incorporate the full set of constraints used in the numerical analysis. The sum rules for \( g_2 \), \( g_3 \), and \( n_4 \) in \( d = 5 \) dimensions are given by

\begin{align}
\texttt{g2sum5d} &= \sum_{j = 0}^{J,\; \text{step}=2} \sum_{i=0}^{N-1} 
\frac{128 (1 + j)^2 (i + 1)^{3/2}}{N^{5/2}}\, \rho_{j, i}, \\
\texttt{g3sum5d} &= \sum_{j = 0}^{J,\; \text{step}=2} \sum_{i=0}^{N-1} 
\frac{128 (1 + j)^2 \left(3 - \frac{4j(2 + j)}{3} \right) (i + 1)^{5/2}}{N^{7/2}}\, \rho_{j, i}, \\
\texttt{n4sum5d} &= \sum_{j = 0}^{J,\; \text{step}=2} \sum_{i=0}^{N-1} 
\frac{128 (1 + j)^2 \left( \frac{4 j (2 + j) (-21 + 4j + 2j^2)}{15} \right) (i + 1)^{7/2}}{N^{9/2}}\, \rho_{j, i}.
\end{align}

We then impose the sum rule relations on the Wilson coefficients:

\begin{align}
\texttt{g2sum5d} &= g_2 - \frac{g_2\, (119 g_2 + 19 g_3)}{2 \times 14336\, \pi^2}, \\
\texttt{g3sum5d} &= g_3 - \frac{749\, g_2^2}{36864\, \pi^2} - \frac{187\, g_2\, g_3}{61440\, \pi^2}, \\
\texttt{n4sum5d} &= \frac{g_2\, (3 g_2 - g_3)}{2304\, \pi^2}.
\end{align}

The sum rules for \( g_2 \), \( g_3 \), and \( n_4 \) in \( d = 6 \) dimensions are given by:

\begin{align}
\texttt{g2sum6d} &= \sum_{j = 0}^{J,\; \text{step}=2} \sum_{i=0}^{N-1} \frac{64 \pi (j+1) (j+2) (2j+3) (i+1)^2}{N^3} \rho_{j, i}, \\
\texttt{g3sum6d} &= \sum_{j = 0}^{J,\; \text{step}=2} \sum_{i=0}^{N-1} 
\frac{64 \pi   (j+1) (j+2) (2 j+3) (j (j+3)-3)(i+1)^3}{N^4} \rho_{j, i}, \\
\texttt{n4sum6d} &= \sum_{j = 0}^{J,\; \text{step}=2} \sum_{i=0}^{N-1} 
\frac{64 \pi   (j+1) (j+2) (2 j+3) (j (j+3) (j (j+3)-13))(i+1)^4}{3\ N^5} \rho_{j, i}.
\end{align}

The relevant relations corrected by one loop take the following form:

\begin{align}
\texttt{g2sum6d} &= \frac{19 g_{2}^2 }{3840 \pi^3}
+ g_{3} \left( \frac{g_{2}}{1152 \pi^3} - 1 \right), \\
\texttt{g3sum6d} &= g_{2} - \frac{g_{2}  \left(60 g_{2} + 11 g_{3}  \right)}{2 \times 26880 \pi^3}, \\
\texttt{n4sum6d} &= \frac{g_2 (2 g_2-g_3 )}{13440 \pi ^3}.
\end{align}

The allowed region for \( g_2 \) and \( g_3 \) is determined by fixing values of these couplings and then searching for a feasible solution \( \rho_{j,i} \) satisfying the above relations. This discretization reduces the problem to a finite-dimensional linear system, which can be analyzed or optimized using standard numerical techniques.

Finally, we fix \( N = 200 \) and \( J = 38 \) in the numerical implementation. This choice yields remarkably stable results: we checked that further increase in numerical resolution and number of spins do not significantly affect the outcome but substantially increase the computational cost. We present our final results below:

\begin{figure}[H]
    \centering
    \begin{subfigure}{0.43\textwidth}
        \centering
        \includegraphics[width=\textwidth]{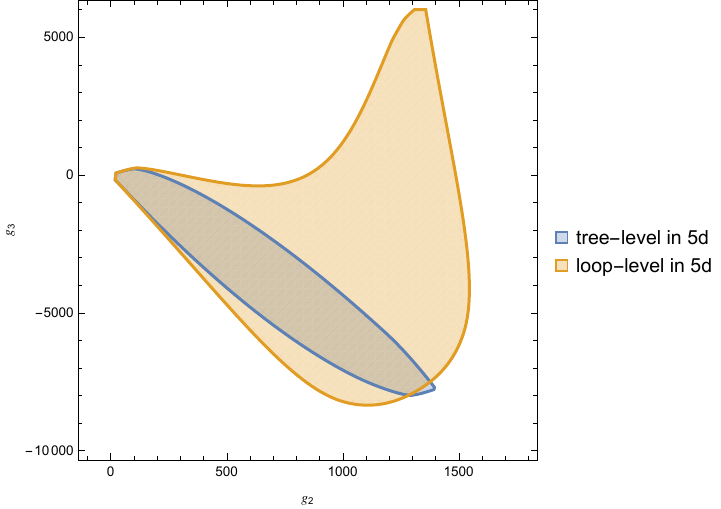}
        \caption{Numerical bounds on $g_2$ and $g_3$ in 5D.}
        \label{fig:Numerical bounds on $g_2$ and $g_3$ in d=5.}
    \end{subfigure}
    \quad
    \begin{subfigure}{0.45\textwidth}
        \centering
        \includegraphics[width=\textwidth]{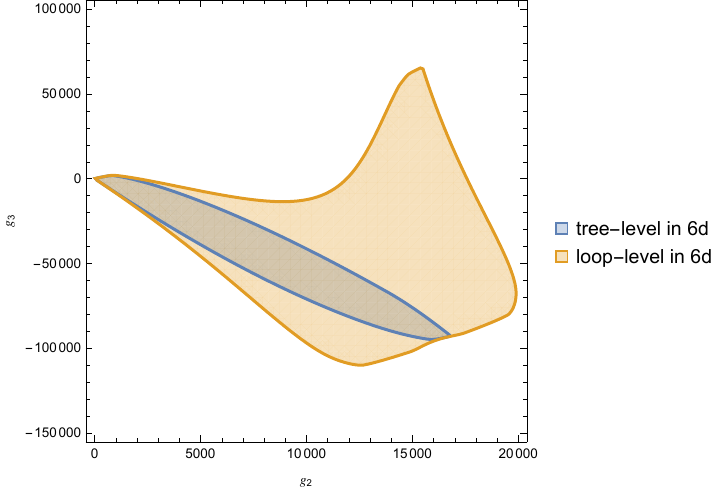}
        \caption{Numerical bounds on $g_2$ and $g_3$ in 6D.}
        \label{fig:Numerical bounds on $g_2$ and $g_3$ in 6D.}
    \end{subfigure}
    \caption{Numerical bounds on $g_2$ and $g_3$ in 5D and 6D.}
    \label{fig:g2-g3-bounds}
\end{figure} 

We observe that when both \( g_2 \) and \( g_3 \) are small, the tree-level and loop-level bounds coincide. This is expected, as loop corrections are negligible in the weak-coupling regime. However, as the couplings increase, loop effects become more significant, leading to dramatic changes in the shape of the allowed region. 

In particular, the upper bound on \( g_2 \) shifts slightly outward, while the bound on \( g_3 \) changes more substantially. Notably, the allowed region is no longer convex. This non-convexity arises because the inclusion of loop effects makes the map between the sum rules and the Wilson coefficients nonlinear. Consequently, even if the geometry of the parameter space defined by positivity remains convex, a nonlinear projection can result in a non-convex image.

\section{Bounds on $g_3/g_2-g_4/g_2$}
\label{sec:g3g4}
In this Section, we find the allowed region for dimensionless quantities $\tilde{g}_3=\frac{g_3}{g_2}$ and $\tilde{g}_4=\frac{g_4}{g_2}$ beyond tree level. The most optimal bound at tree level for $\tilde{g}_3$ and $\tilde{g}_4$ was first obtained numerically~\cite{Caron-Huot:2020cmc} and then analytically reproduced by~\cite{Chiang:2021ziz} as a solution to the corresponding moment problem. In this section, we will use the analytic method called moment problem and also the numeric method of linear program introduced in previous section. From the point of view of $a$-geometry, we use exactly the same analytic methods as in~\cite{Chiang:2021ziz}, truncated up to $k=5$ order. We reproduce previous tree level bounds on $\tilde{g}_3$ and $\tilde{g}_4$ in 4D, 5D, 6D, and then incorporate the loop effect in 5D and 6D. The loop effects in 4D is more subtle as can be seen from the modified null constraints, we will present only the weak coupling limit bounds in 4D. 

\subsection{Constraints in 5D and 6D \label{sec5.1}}

For the analytic method, the conditions that $\tilde{g}_3$ and $\tilde{g}_4$ should satisfy are listed in the Appendix~\ref{appc}. We scan the parameter space of $\tilde{g}_3$ and $\tilde{g}_4$ and include only points that satisfy these conditions. 

For the numerical method, we discretize the integral of the moments into summation and then impose only the unitarity condition and modified null constraints, i.e. Eq.~\eqref{d5null} in 5D and Eq.~\eqref{d6null} in 6D. The truncation of the angular momentum $J$ is 38 and the step $N$ is 200 as in previous section.

\begin{figure}[htp]
    \centering
    \begin{subfigure}{0.43\textwidth}
        \centering
        \includegraphics[width=\textwidth]{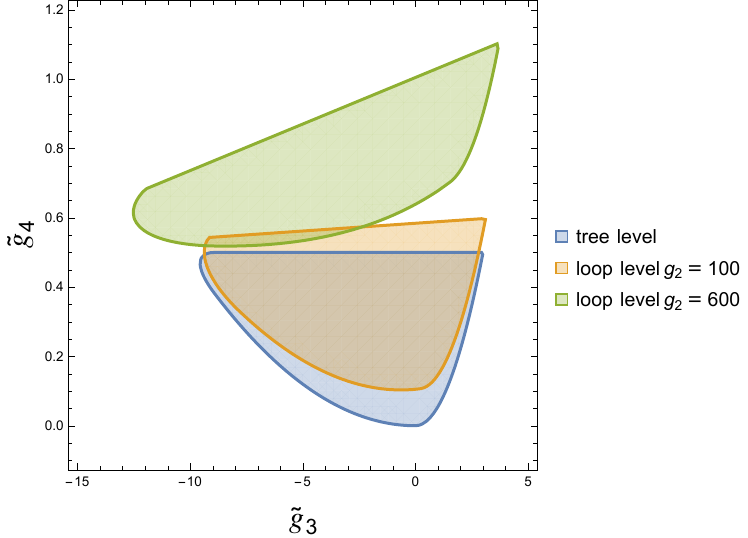}
        \label{fig5a}
    \end{subfigure}
    \quad
    \begin{subfigure}{0.45\textwidth}
        \centering
        \includegraphics[width=\textwidth]{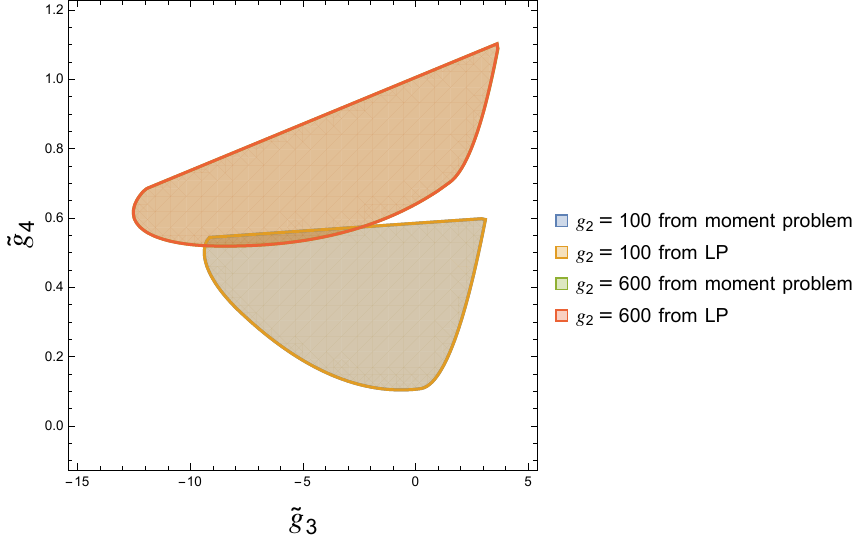}
        \label{fig5b}
    \end{subfigure}
    \caption{(a) Allowed region for $\tilde{g}_3$-$\tilde{g}_4$ with only positivity included, the blue shaded region denotes tree level bounds, the yellow and green shaded region denote the loop corrected bounds for $g_2=100$ and $g_2=600$ respectively. (b) Loop corrected bounds for $\tilde{g}_3$-$\tilde{g}_4$ for different values of $g_2$. The bounds are calculated using moment problem technique and linear program (LP) technique described above, the bounds derived using two different methods matches with almost no differences.}
    \label{g3g4}
\end{figure} 

The result in 5D is shown in Fig.~\ref{g3g4}. In the left panel of Fig.~\ref{g3g4}, the allowed region is derived by using only positivity, i.e. $\rho_j\geq0$. As $g_2$ increases, the allowed region starts to deviate from the tree level one. Interestingly, when $g_2$ becomes larger than 600, there will not be any intersection with the tree level one. We plotted loop level bounds using the analytic and numerical methods in the right panel of Fig.~\ref{g3g4}, where "LP" stands for linear programming method. The result from  these two methods match almost exactly, showing that the numerical method is convergent for our choice of $N$ and $J$.

\begin{figure}[htp]
    \centering
    \includegraphics[width=0.5\linewidth]{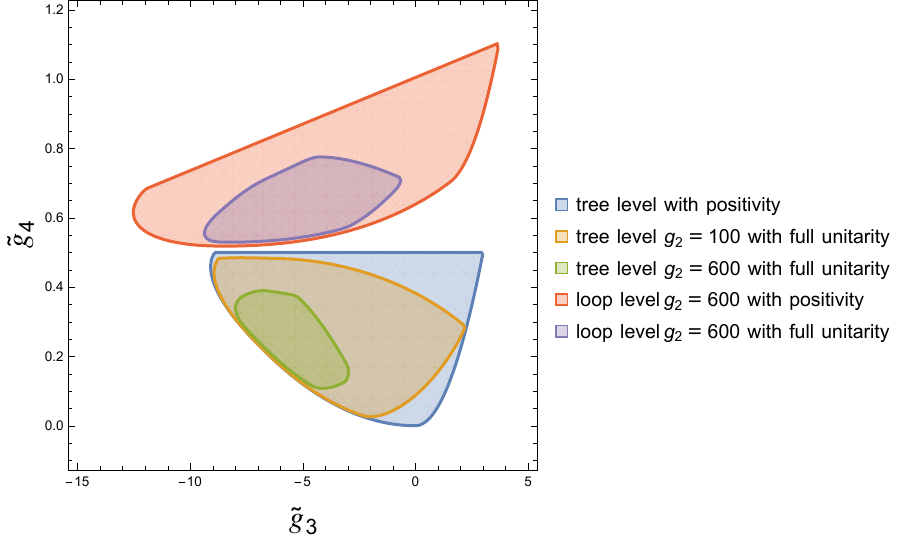}
    \caption{Comparison of positivity bounds and full unitarity bounds.}
    \label{PvsFU5d}
\end{figure}

In the method of linear programming, the full unitarity condition, i.e. $0\leq\rho_j\leq2 $, is easy to implement, while it is more difficult to implement such a condition analytically. Given that the numerical result from linear program matches well with the analytical one including only positivity, we assume that it also converges quickly when full unitarity is included. The result is shown in Fig.~\ref{PvsFU5d}. As one would expect, the bounds with full unitarity are stronger than with only positivity. The full unitarity leads to non-linearity like including loop effect did, therefore, the projective $g_3/g_2-g_4/g_2$ bounds from positivity only now becomes non-projective, namely, they depend on the value of $g_2$ we take.

\subsection{Bounds in 4D in a weak coupling limit}

\begin{figure}[htp]
    \centering
    \includegraphics[width=0.5\linewidth]{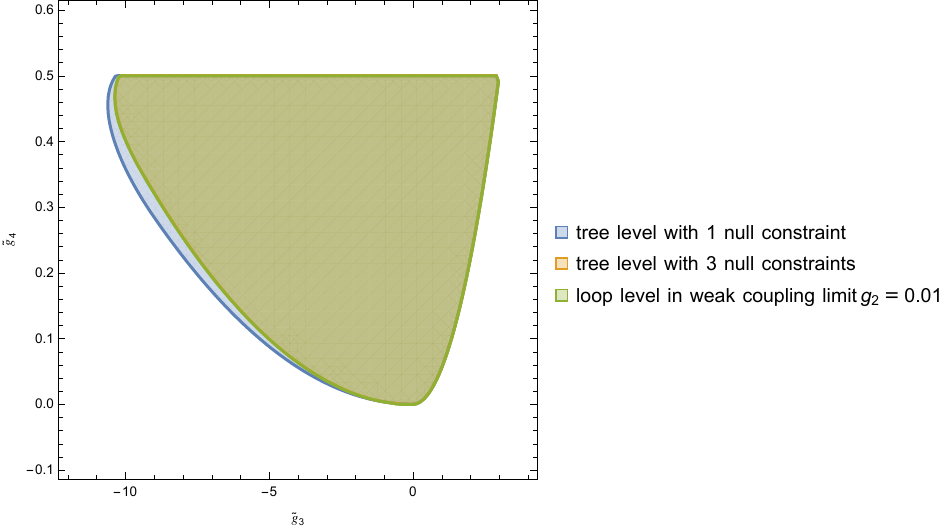}
    \caption{$\tilde{g}_3-\tilde{g}_4$ bounds in 4D in weak coupling limit. Tree level with 1 null constraint is analytic result from~\cite{Chiang:2021ziz}, the rest region is plotted using numerical linear program method.}
    \label{g3g44d}
\end{figure}

In 4D, we can not use directly the $n_4$ null constraint due to the presence of $\log(-t)$. But there is also a $\log(-t)$ coming from $n_5$, taking combinations of $n_4$ and $n_5$ to eliminate the $\log(-t)$ divergence, we get a usable null constraint $n_{45}$ (see Eq.~\eqref{n45}). However, we found it is still not enough to have a lower bound on $\tilde{g}_3$ using the numerical linear programming method. In fact, we need one at least one more constraint which could be obtained from combinations of $n_6$ and $n_4$ (see Eq.~\eqref{n46}). With these two conditions and positivity condition, we reproduced the tree level $\tilde{g}_3-\tilde{g}_4$ bound in 4D in weak coupling limit of our loop-level bound, i.e. take $g_2=0.01$. The result is shown in Fig.~\ref{g3g44d}.

 The blue-shaded region is an analytic result from moment problem~\cite{Chiang:2021ziz} with the use of one tree-level null constraint $n_4=0$. The green-shaded region is numerical result from with the use of tree-level mapping from a-geometry to g-geometry and three null constraints, $n_4=0$, $n_5=0$, $n_6=0$. The yellow-shaded region is the numerical result from linear program with the use of loop level mapping from a-geometry to g-geometry (see Eq.~(\ref{4dnull}--\ref{d6g4})) and the three combined null constraints. The green region and yellow region matches, showing that our loop level null constraints can have a smooth limit to tree level one. 

\begin{figure}[htp]
    \centering
    \includegraphics[width=0.4\linewidth]{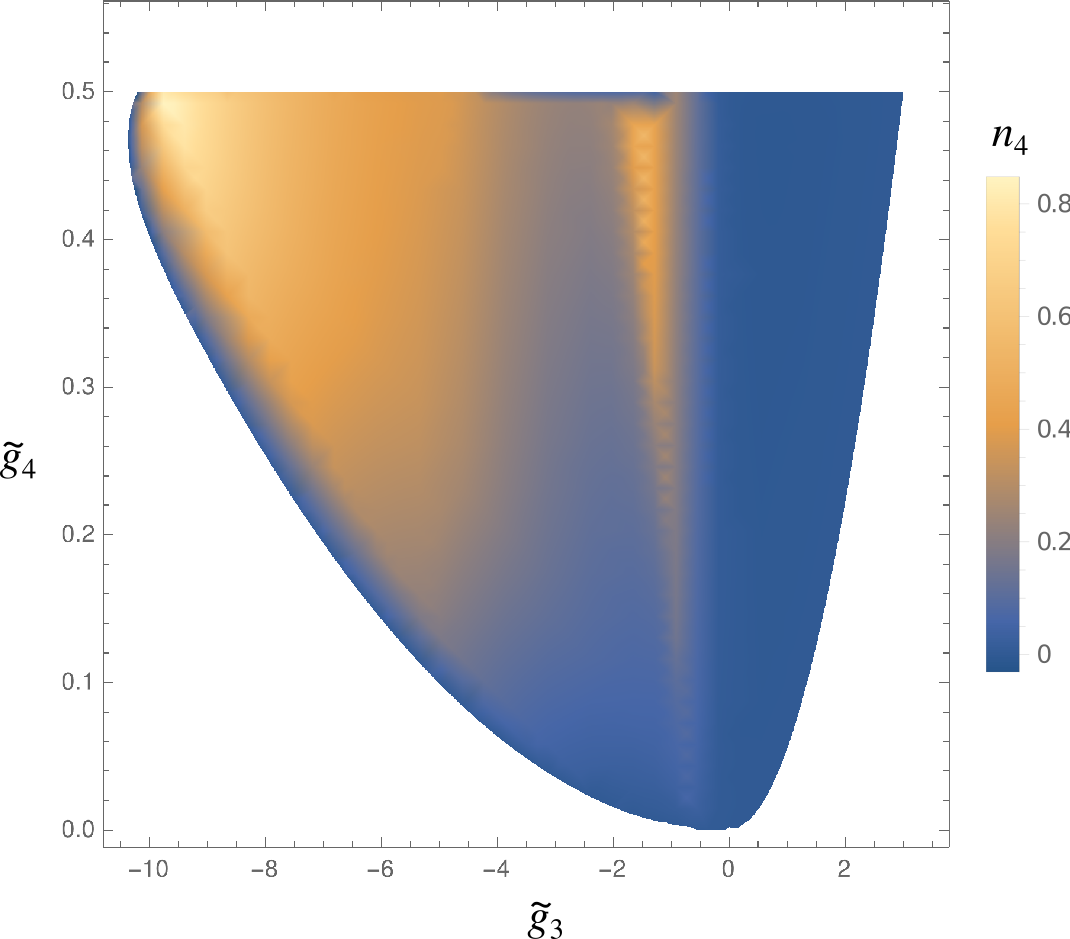}
    \caption{The maximal values of $(-\tilde{g}_3) n_4$ in $\tilde{g}_3-\tilde{g}_4$ bounds. On the left boundary corresponding to the lower bound on $\tilde{g}_3$ the value of $n_4$ is zero.}
    \label{3dplot}
\end{figure}

 It is surprising that our two combined null constraint conditions could give the same result as the one from the use of three tree level null constraints. We could see this from numerical result for the maximum values of $n_4$, as shown in the Fig.~\ref{3dplot}. We noticed that the null constraints are exactly zero on the left boundary while in the middle of the figure it's value may not be zero, although it is still bounded from above. Recall that this boundedness  doesn't follow from an attempt to match the null constraint condition with logarithmically divergent forward limit of $n_4$ computed from EFT. There could be a lower bound on $n_4$ but the upper bound diverges. However, we found it is restored in a non-trivial way\footnote{There is no contradiction between the upper bound on $n_4$ and $\log{t}$ divergence of the corresponding expression computed in EFT. This situation just illustrates the breaking of the relation between EFT expressions and moments when the forward limit is divergent (see also Appendix \ref{sec:B} for more detailed discussion).}. 

Here we provide a heuristic  explanation for such a behavior of the $n=4$ null constraint. It is easy to understand the origin of the right boundary of $\tilde{g}_3$ because $\tilde{g}_3=\left<(3-\mathcal{J}^2)/\mu^4\right>\leq\left<3/\mu^4\right>$, the boundary is given by such $\rho_j(\mu)$ that it vanishes for $j\ne0$. As for the left boundary, from the EFThedron or moment problem~\cite{Arkani-Hamed:2021ajd,Chiang:2021ziz}, we know that it is determined by saturating the so called "polytope boundary" (see Eq.~\eqref{form:cons4} in Appendix.~\ref{sec:C})
\begin{equation}
     \left<\frac{(\mathcal{J}^2-6)(\mathcal{J}^2-20)}{\mu^{5}}\right>= 0,\quad \mathcal{J}^2=j(j+1)\quad \text{with $j\in\{0,2,4,6...\}$}.   
\end{equation}
This gives us non-trivial information about the theories living on that boundary --- it can contain only spin-2 and spin-4 contributions in the density of states. Indeed, the saturation of this condition can happen only if the other $\rho_j=0$ because they can give only positive contributions.

Recall the loop-level null constraints in the weak coupling limit are
\begin{gather}
     18\tilde{g}_3 \left<\frac{\mathcal{J}^2(\mathcal{J}^2-8)}{\mu^5}\right>+\left <\frac{{\mathcal J}^2 \left[ {\mathcal J}^2 \left( 2 {\mathcal J}^2 - 43 \right) + 150 \right]}{\mu^6} \right>=0  ,\label{450}\\
     -\frac{7}{6}\left<\frac{\mathcal{J}^8-44\mathcal{J}^6+588\mathcal{J}^4-2448\mathcal{J}^2}{\mu^{7}}\right>+6(7\tilde{g}_3^2+24\tilde{g}_4)\left<\frac{\mathcal{J}^2(\mathcal{J}^2-8)}{\mu^5}\right>=0. \label{460}
\end{gather}

Remember that the bracket means summation over $j$ and integration over $\mu$ with the positive measure $\rho_j(\mu)$ and some normalization factor (see Eq.~\eqref{brackets}). Knowing that only $j=2$ and $j=4$ contribute, we can separate the each term in the above equation into two terms
\begin{gather}
18\tilde{g}_3\left(\left<\frac{-12}{\mu^{5}}\right>_{\rm{j=2}}+\left<\frac{240}{\mu^{5}}\right>_{\rm{j=4}}\right)+\left<\frac{-216}{\mu^6}\right>_{\rm{j=2}}+\left<\frac{1800}{\mu^6}\right>_{\rm{j=4}}=0. \label{pw45}\\
     \left<\frac{2016}{\mu^{7}}\right>_{\rm{j=2}}+\left<\frac{6720}{\mu^{7}}\right>_{\rm{j=4}}+6(7\tilde{g}_3^2+24\tilde{g}_4)\left(\left<\frac{-12}{\mu^5}\right>_{\rm{j=2}}+\left<\frac{240}{\mu^5}\right>_{\rm{j=4}}\right)=0. \label{pw46}
\end{gather}
where $\left<\cdots\right>_{\rm{j=2}}$ means $n_2^{(4)} \int_{1}^\infty  {\rm d}\mu \, \rho_2(\mu) \cdots$ with $n_2^{(d)}$ the normalization factor defined in Eq.~\eqref{nf}. $\left<\cdots\right>_{\rm{j=4}}$ is similarly defined by changing $j=2$ to $j=4$. 

Next, as we can learn from the numerical result in Fig.~\ref{g3g44d}, on the boundary, the $\tilde{g}_3^2$ is much larger than $\tilde{g}_4$, so it is reasonable to neglect the $\tilde{g}_4$ term since we are trying to show that on the boundary of $\tilde{g}_3-\tilde{g}_4$ bounds, the null constraints Eq.~\eqref{450}, Eq.~\eqref{460} are the same as the tree level null constraints $n_4,n_5,n_6=0$.


Now we do a rescaling of $\mu$, define $\mu\rightarrow\mu^r=\sqrt{42}|\tilde{g}_3|\mu$ and a new bracket $\left<...\right>^r$ such that it is the same as the original bracket $\left<...\right>$ except for integrate region starts from $\sqrt{42}|\tilde{g}_3|$ to infinity and the spectral becomes $\rho_j(\mu^r/\sqrt{42}|\tilde{g}_3|)$ now. After this rescaling, $\tilde{g}_3$ is absorbed into $n_4$ in both finite combinations of null constraints. Thus, Eq.~\eqref{pw46} becomes
\begin{gather}
     \left<\frac{2016}{(\mu^{r})^{7}}\right>_{\rm{j=2}}^r+\left<\frac{6720}{(\mu^{r})^{7}}\right>^r_{\rm{j=4}}+\left<\frac{-12}{(\mu^{r})^{5}}\right>^r_{\rm{j=2}}+\left<\frac{240}{(\mu^{r})^{5}}\right>^r_{\rm{j=4}}=0. \label{460r}
\end{gather}
Notice that only the third term is negative in Eq.~\eqref{460r}, so the equality is possible only when $\rho_j(\mu^r/\sqrt{42}|\tilde{g}_3|)$ has support at large $\mu^r$ (of order $\sim O(10^1)$).


Using the same argument for Eq.~\eqref{pw45}, after the rescaling we obtain
\begin{gather}
     \left(\left<\frac{-1}{(\mu^{r})^{5}}\right>_{\rm{j=2}}^r+\left<\frac{20}{(\mu^{r})^{5}}\right>^r_{\rm{j=4}}\right)+\left(\left<\frac{-6.47}{(\mu^{r})^{6}}\right>^r_{\rm{j=2}}+\left<\frac{53.9}{(\mu^{r})^{6}}\right>^r_{\rm{j=4}}\right)=0.
\end{gather}
Since the numerators of both terms are of the same order $\sim O(10^1)$ while we have see that $\rho_j$ is non-zero for vary large $\mu$, the terms corresponding to the $n_5$ null constraint (the second bracket) are suppressed compared to the $n_4$ combination by the additional $\mu^r$ in denominator. Thus, $n_4$ has to be approximately zero separately from $n_5$. In this case, all other terms in the null constraints are small, so the null constraints $n_4, n_5, n_6$ will be separately close to zero on the 'polytope boundary'.

\section{Conclusions}
\label{sec:conclusions}
In this paper, we considered a 4-particle amplitude with loop contributions describing the scattering of massless scalars with shift symmetry. We formulated connections between Wilson coefficients in the IR theory with the branch cut integrals (brackets) in the UV as it was done in previous positivity bounds works. In the process of constructing these relations, we need to cancel the IR divergences brought in by massless loop contributions. We found that loop corrections not only introduce the scale dependence of the EFT coefficients, but also change the null constraints which comes from $s,t,u$ full crossing symmetry. We found that in spacetime dimensions higher than four this change is simple. The null constraints are no longer "null", but equal to a constant depending on the $\beta$-functions. This way, in the weak coupling limit the usual null-constraints are restored. The impact of this modification of null constraint is profound, as it brings many new features that were not observed previously. Firstly, the modified null constraint itself is nonlinear in Wilson coefficients. From geometric point of view it means that this constraint is represented by the curved surface intersecting the allowed space of the moments (a-geometry of EFThedron). The resulting constraints, especially obtained away from the weak coupling limit may not be shown as convex areas, which used to be the case for tree-level bounds. 

We have studied the upper bounds on the EFT coefficients $g_3$ and $g_2$ in five and six spacetime dimensions and found that their tree-level constraints obtained along the lines of \cite{Caron-Huot:2020cmc,Chiang:2022ltp} are significantly modified. In addition, the relation between the moments and EFT couplings is also non-linearly corrected, and these corrections become large. For this reason, the upper bounds obtained in this work rely on assumption that the higher-derivative terms in EFT are suppressed. This assumption can be avoided by taking the arc radius in the dispersion relations to be a bit smaller. This choice would provide a more stable bound, however, it would be non-optimal. A proper model-independent optimization of these constraints would require to resum all one-loop corrections to EFT coefficients which may be not possible.

Contrary to the unstable behavior of the constraints on the absolute value of $g_2$ we observed that the constraints on ratios $g_3/g_2$ and $g_4/g_2$ received small corrections at loop level. The inclusion of full unitarity condition leads to stronger constraints, especially for large $g_2$. These constraints obtained numerically are reproduced analytically from moments problem with very high precision even for large values of $g_2$. Remarkably, we found that inclusion of full unitarity condition selects the parameters of $g_3/g_2$ and $g_4/g_2$ close to the boundary corresponding to the 'extremal' EFTs \cite{Caron-Huot:2020cmc} when $g_2$ is large.

We found also that the forward limit expansion of the amplitudes in four dimesions makes this case very different from higher dimensions. In four dimensions all null constraints receive an unbounded correction with $\log{(-t)}$ term which invalidates the direct use of tree-level approaches. However, we found that one can construct linear combinations of null-constraints which are finite in forward limit. In this case we can relate them to the corresponding moments, as it is argued in detail in the Appendix B. For reproducing the weak coupling limit we found that it is enough to take two finite combinations of $n_4$, $n_5$ and $n_6$ as constraints in our numerical procedure. In this way, the same boundary as at tree-level is reproduced with high precision. This actually means that even if we are not requiring $n_4=0$ this condition can be still obtained on the relevant boundary. It shows that for small $g_2$ the analytically obtained bound for $g_3/g_2$ is very close to the most optimal one even if loop corrections are included.

The constraints in four dimensions in the strong coupling limit would require at least two-loop corrections to be included, as they would modify $n_6$ null constraint by extra $\log^2$ term. For this reason, we do not discuss here the bounds on the absolute value of $g_2$ leaving this analysis for future studies.

The results and methods developed in this work can be straightforwardly applied to constraining more complicated EFTs including particles with spin and gravity. More rigorous bounds beyond weak coupling assumptions require knowledge about the structure of the IR loop corrections. A systematic way of obtaining the bounds includes the formulation of null constraints at the loop level which depend on the concrete EFT, its field content, and couplings. It seems to be particularly interesting to obtain consistency conditions on $\beta$-functions of the theory which encode information about the number of states running in loops and their spins. 

It is a very interesting problem for future studies, whether the same methods based on defining quantities that are finite in $t\rightarrow 0$ limit can be extended and applied to theories including gravity. The main difference is caused by the presence of the graviton pole which makes it more tricky. In addition, the convergence properties of the moment sums and integrals have to be separately checked. Getting finite moment integrals may require additional smearing over $t$. Very recent papers \cite{Chang:2025cxc,Beadle:2025cdx} show how the finite $t$ bounds with gravity can be obtained numerically. It seems to be an interesting question whether these bounds can be understood analytically from the moment problem. In addition, the role of the full unitarity condition which should improve the bounds is also not yet clear in the existing results. Formulation of the bounds with gravity as $L$-moment problem in a mathematically consistent way seems to be a promising direction for future study.

\subsection*{Acknowledgements}

Authors are indebted to Shuang-Yong Zhou, Tong Arthur Wu and Shilin Wan for several enlightening conversations. A.~T. is also grateful to Laurentiu Rodina for his hospitality at BIMSA (Beijing) and for illuminating discussions on the EFT-hedron methods. A.~T. thanks Brando Bellazzini, Kelian Haring, Denis Karateev, Francesco Riva, Alexey Koshelev and Alexander Zhiboedov for valuable comments, feedback, and criticism. The work of A.~T. was supported by the National Natural Science Foundation of China (NSFC) under Grant No. 1234710.

\appendix
\section{Smeared dispersion relations.}
\label{sec:A}
The assumption of the boundedness of the amplitudes by $s^2$ is valid only for negative values of $t$. Even in the case of massive EFTs the Martin-Froissart bound can be proven only for negative $t$, while there is no much information about the behavior of the amplitudes for unphysical $t>0$ values. For this reason, the dispersion relations used in the main text of this paper are well-defined only for $t<0$, and due to their non-analytic behavior at $t\rightarrow 0$ one has to define them and their $t$-derivatives in a more rigorous way, i. e. in the distributional sense.

The arc integrals $M_n$ can be integrated with respect to $t$ with the smooth functions,
\begin{equation}
\label{smearing}
    \bar{M}_n(q_0)=\int_0^{q_0} dq f(q)M_n(q).
\end{equation}
Here we take $t=-q^2$. The basis for such functions $f(q)$ can be selected as
\begin{equation}
    f(q)=\frac{q^{\gamma}(q_0-q)^{\alpha}}{q_0^{\gamma+\alpha+1}}\frac{\Gamma(2+\alpha+\gamma)}{\Gamma(1+\alpha)\Gamma(1+\gamma)}.
\end{equation}
Here $\gamma$ and $\alpha$ have to be chosen such that all integrals converge. This operation allows to define the dispersion relations as functionals on the space of the functions $f(q)$. $t$-derivatives of the dispersion relations can be defined as the integrals with $(-1)^k \partial^k f(q)/(\partial q^2)^k$. 

The described 'smearing' operation allows also to guarantee the convergence of the partial wave sums after the integration with $f(q)$. Recall that in 4 dimensions partial wave sums cannot be proven to be convergent at small finite $t$ if one assumes only the unitarity condition. To show how the situation is improved by smearing we can take $M_2$ and use that 
\begin{equation}
 G=\int_0^{q_0} d q f(q)\, P_J\left(1-\frac{2q^2}{\mu }\right)=  \, _4F_3\left(-j,j+1,\frac{\gamma }{2}+\frac{1}{2},\frac{\gamma }{2}+1;1,\frac{\alpha
   }{2}+\frac{\gamma }{2}+1,\frac{\alpha }{2}+\frac{\gamma
   }{2}+\frac{3}{2};\frac{q_0^2}{\mu }\right).
\end{equation}
The large $J$ expansion of the hypergeometric function $_4F_3$ contains oscillating and non-oscillating parts,
\begin{equation}
\begin{split}
     G&=\frac{2^{-\alpha -\gamma -2} \sin (\pi  \gamma ) \Gamma (\alpha
   +\gamma +2) z^{\frac{1}{2} (-\alpha -\gamma -2)}}{\pi ^{3/2} \Gamma (\alpha +1)}
   \left(2^{\alpha } \alpha  \Gamma \left(\frac{1}{2}-\frac{\gamma
   }{2}\right) \Gamma \left(\frac{\gamma }{2}+1\right) z^{\alpha
   /2}+\right.\\
   &+2^{\gamma +1} \Gamma (\alpha +1) \Gamma (-\gamma )
   z^{\frac{\gamma }{2}+\frac{1}{4}} \sin \left(\frac{1}{4}
   \left(2 \pi  \alpha -8 \sqrt{z}+\pi \right)\right)-\left.2^{\alpha }
   \Gamma \left(-\frac{\gamma }{2}\right) \Gamma
   \left(\frac{\gamma +1}{2}\right) z^{\frac{\alpha
   +1}{2}}\right).
   \end{split}
\end{equation}
Here 
\begin{equation}
    z=\frac{(2 j+1)^2 q_0^2}{\mu }.
\end{equation}
 One can require oscillating term to be suppressed which can always be achieved by taking large enough $\alpha$. Thus, the sum in the right-hand side of dispersion relations in the large $J$ limit has the form
 \begin{equation}
     \sum_J (2j+1) G \,{\rm Im\,} f_j \propto \sum_j j^{-\gamma }\,{\rm Im\,} f_j<\sum_j 2\,j^{-\gamma}.
 \end{equation}
This sum is convergent for $\gamma>1$, hence the dispersion relations (without $t$-derivatives) are well-defined under the smearing operation provided this condition is satisfied. Dispersion relations with $t$-derivatives can be defined in a similar way, however, they would require higher powers of $\gamma$ for convergence of the partial wave sums.

It is important to note here that it could also happen that a particular low-energy EFT imposes stronger constraints on the dependence of ${\rm Im\,}f_j(s)$ in the UV that just unitarity. Indeed, the condition $\gamma>1$ means that the $1/t$ singularity in the IR can be integrated with such a smearing function and lead to a finite result in both sides of the dispersion relation even if the unitarity condition is saturated. However, if we do not consider graviton exchange contribution the left hand side is finite in the forward limit, which means the smearing function with $\gamma>-1$ can also be used. It implied that the right hand side of this dispersion relation should also be finite for $-1<\gamma<1$. It doesn't follow from the unitarity condition ${\rm Im}\,f_j<2$. Instead, it encodes more specific information on the ${\rm Im}\,f_j$ in the UV, such that the pole singularity is not produced in the IR. In general, such information can be extracted in a similar way to the relations obtained in \cite{Haring:2024wyz} which guarantee the reproduction of the graviton pole from the dispersion relation. In a theory without gravity one needs to require the reproduction of the constant term instead of a pole. To start with, this corresponds to the fact that the right-hand side of the dispersion relation must be convergent for any $\gamma>-1$. More precise requirements on the UV theory imposed by the specific form of IR singularities in the forward limit can also be extracted. As for the purposes of the current work it is enough to point out the convergence of the right-hand side of the dispersion relations corresponding to the finite forward limit, we leave exploration of these relations for future study.

The other advantage of the smearing operation is that, given the proper choice of $\alpha$ and $\gamma$, one can get the set of positive-definite functions after integration of $f(q)$ with Legendre polynomials. For the reason of complexity of the resulting functions, it seems to be not possible to obtain any analytic results based on the moments problem for large $q_0$ (i.e. beyond the forward limit). However, for small $q_0\ll \Lambda$ one can still use the series expansion of the smeared dispersion relations. This expansion returns us back to $t\rightarrow 0$ expansions used in the paper, while the rigorous definitions for the dispersion relations and their derivatives can be now given in the distributional sense. In addition, smearing also allows to find the conditions under which the usual relation between EFT couplings and moments at $t\rightarrow 0$ can still be used.

\section{On the connection between moments and EFT parameters in the presence of forward limit singularities.}
\label{sec:B}

For EFTs without mass gap it is not obvious whether the $t\rightarrow 0$ limit can be used for building the link between EFT couplings and moments of the density of states obtained as the expansion of the dispersion relations in series around $t\rightarrow 0$. This link is obstructed by the presence of non-analytic terms in arc integrals $M_n$ and their $t-$derivatives. Let us illustrate this link using the example of $n=4$ null constraint in four dimensions. 

\begin{equation}
\begin{split}
n_4=&\frac{\partial^2 M_2}{\partial t^2}- 6 M_4=2 b_2 \log (\epsilon^2)-2b_2\log (-t)-3 b_2+2 c_2\epsilon^2+\\
+&t \left(\frac{30 b_1}{\epsilon ^2}+\frac{6 b_2}{\epsilon ^2}-6 c_2 \log
   (-t)-6 \left(5 c_1+c_2\right) \log \left(\epsilon ^2\right)-5 c_2-12
   g_5\right)+ O(t^2),
\end{split}
\label{B-4dnull}
\end{equation}

The presence of several non-analytic terms, and, moreover, the divergence of the forward limit of $n_4$ leads to the obstruction that one can still use the moments expansions of the right-hand side of the corresponding dispersion relations. However, one can address the question whether the dispersion relation could still make sense for small but finite $t<0$ after smearing.

If we apply the smearing operation \eqref{smearing} to the expression
\begin{equation}
    M=A+ B \log{(-t)}+ C t \log{(-t)}+D t,
\end{equation}
we get
\begin{equation}
\begin{split}
    \bar{M}=&A+2 B \left(-H_{\alpha +\gamma +1}+H_{\gamma }+\log
   \left(q_0\right)\right)+C\,\frac{2 (\alpha +1)  q_0^2
   \left(\alpha  (2 \gamma +3)+2 (\gamma +2)^2\right)}{(\alpha
   +\gamma +2)^2 (\alpha +\gamma +3)^2}+\\
   +&D\,\frac{(\gamma +1) (\gamma
   +2)  q_0^2}{(\alpha +\gamma +2) (\alpha +\gamma +3)}+C\,\frac{2 (\gamma +1) (\gamma +2)  q_0^2 \left(-H_{\alpha
   +\gamma +1}+H_{\gamma }+\log \left(q_0\right)\right)}{(\alpha
   +\gamma +2) (\alpha +\gamma +3)},
   \end{split}
\end{equation}
where $H_n$ is a harmonic number, and for $\gamma\rightarrow-1$ at large $\alpha$ we have
\begin{equation}
    H_{\gamma }-H_{\alpha +\gamma +1}=-\log (\alpha )-\frac{1}{\gamma +1}-\gamma_E+\left(\frac{\pi
   ^2}{6}-\frac{1}{\alpha }\right) (\gamma +1)-\frac{1}{2 \alpha
   }.
\end{equation}
Here $\gamma_E\approx 0.5772$ is an Euler-Mascheroni constant. 

The corresponding right-hand side of the dispersion relation (before summation with respect to $J$ and integration over $\mu$) can be written as
\begin{equation}
    F_{n4}=\frac{\partial^2 F_2}{(\partial(q^2))^2}-6 F_4,
\end{equation}
where we take
\begin{equation}
F_2=\left(\frac{2}{\mu ^3}+\frac{6 q^4}{\mu ^5}+\frac{3 q^2}{\mu
   ^4}\right) \, _2F_1\left(-j,j+1;1;\frac{q^2}{\mu }\right),
\end{equation}
\begin{equation}
F_4=\frac{2}{\mu ^5}\, \, _2F_1\left(-j,j+1;1;\frac{q^2}{\mu }\right),
\end{equation}
provided that $q^2\ll \mu$. In this limit (however, we don't assume here $j^2 q^2\ll \mu$, so the expression is also valid for large spins)
\begin{equation}
   F_{n4}= \frac{j (j+1) }{\mu ^5}\left(\left(j^2+j-2\right) \,
   _2F_1\left(2-j,j+3;3;\frac{q^2}{\mu }\right)-6 \,
   _2F_1\left(1-j,j+2;2;\frac{q^2}{\mu }\right)\right).
\end{equation}
After smearing we obtain ($q_0^2\ll\mu$)
\begin{equation}
\begin{split}
   \bar{F}_{n4} &=\frac{\sqrt{\pi } j (j+1) 2^{-\alpha -\gamma } \Gamma (\alpha
   +\gamma +2) }{\mu ^5}\times\\
   &\times\left(\left(j^2+j-2\right) \,
   _4\tilde{F}_3\left(2-j,j+3,\frac{\gamma +1}{2},\frac{\gamma
   +2}{2};3,\frac{1}{2} (\alpha +\gamma +2),\frac{1}{2} (\alpha
   +\gamma +3);\frac{q_0^2}{\mu }\right)-\right.\\
   &\left.-3 \,
   _4\tilde{F}_3\left(1-j,j+2,\frac{\gamma +1}{2},\frac{\gamma
   +2}{2};2,\frac{1}{2} (\alpha +\gamma +2),\frac{1}{2} (\alpha
   +\gamma +3);\frac{q_0^2}{\mu }\right)\right),
   \end{split}
\end{equation}
where $_4\tilde{F}_3$ stands for the regularized hypergeometric function.\footnote{The corresponding function name in Wolfram Mathematica is HypergeometricPFQRegularized[].} The series expansion of this expression at small $q_0$ reads
\begin{equation}
    \begin{split}
     &  \bar{F}_{n4} = \frac{j (j+1) \left(j^2+j-8\right)}{\mu ^5}-\frac{q_0^2
   \left((\gamma +1) (\gamma +2) (j-1) j (j+1) (j+2)
   \left(j^2+j-15\right)\right)}{3 \left(\mu ^6 (\alpha +\gamma
   +2) (\alpha +\gamma +3)\right)}+\\
   &+\frac{(\gamma +1) (\gamma +2)
   (\gamma +3) (\gamma +4) (j-2) (j-1) j (j+1) (j+2) (j+3)
   \left(j^2+j-24\right) q_0^4}{24 \mu ^7 (\alpha +\gamma +2)
   (\alpha +\gamma +3) (\alpha +\gamma +4) (\alpha +\gamma
   +5)}-\\
   &\frac{q_0^6 \left((\gamma +1) (\gamma +2) (\gamma +3)
   (\gamma +4) (\gamma +5) (\gamma +6) j (j^2-1)
   (j^2-4) (j^2-9) (j+4) \left(j^2+j-35\right)\right)}{360 \left(\mu
   ^8 (\alpha +\gamma +2) (\alpha +\gamma +3) (\alpha +\gamma +4)
   (\alpha +\gamma +5) (\alpha +\gamma +6) (\alpha +\gamma
   +7)\right)}\\
   &+O\left(q_0^8\right).
    \end{split}
\end{equation}

This expression unravels the structure of the hypergeometric series typically emerging after smearing of the right-hand side of the dispersion relations. The first term reproduces the familiar combination appearing in the $n=4$ null constraint which can be obtained expanding the Legendre polynomials at $t=0$. A remarkable feature emerging in this series is that all higher-order terms in $q_0$ come with a common multiplier $(\gamma+1)$. Thus, one can expect that these terms can be suppressed by taking small value of $\gamma+1$ even if $q_0$ is kept finite. In the other words, it can be possible to exchange the order of taking limits $q_0\rightarrow 0$ and $\gamma\rightarrow -1$. This is possible if the corresponding sum and integral lead to a finite result,
\begin{equation}
    \bar{M}=\left<\bar{F}_{n4}\right>=\sum_{j=0}^{\infty}\int_{\epsilon^2}^{\infty}d\mu\, 16 \,(2j+1)\,\bar{F}_{n4}\,{\rm Im\,}f_j(\mu).
\end{equation}
In the limit of $\gamma\rightarrow -1$ the left-hand side of the smeared dispersion relation is,
\begin{equation}
\begin{split}
   \bar{M}=& -\frac{2 B}{\gamma +1}+\left(A-2 B \psi ^{(0)}(\alpha +1)+2 B \log
   \left(q_0\right)-2 \gamma_E  B\right)+\\
   +&(\gamma +1) \left(-2 B \psi
   ^{(1)}(\alpha +1)+\frac{\pi ^2 B}{3}+\frac{2 C q_0^2 \log
   \left(q_0\right)}{(\alpha +1) (\alpha +2)}-\frac{2 C
   q_0^2 \psi ^{(0)}(\alpha +1)}{(\alpha +1) (\alpha
   +2)}+\right.\\
   +&\left.C \,\frac{q_0^2 ((\alpha +1)
   (\alpha +2) D-2 (-\alpha  (\alpha +1)+\gamma_E  (\alpha
   +2) (\alpha +1)+1) }{\left(\alpha ^2+3 \alpha
   +2\right)^2}\right)+O\left((\gamma +1)^2\right).
   \end{split}
\end{equation}
Here $\psi^{(0)}$ and $\psi^{(1)}$ are digamma function and its derivative, respectively.
From this expression for $\bar{M}$ one can immediately see the problem of taking the limit $\gamma\rightarrow -1$ in both sides of the dispersion relation, in the presence of the logarithmic IR singularity coming with the constant $B$. In this case the naive identification 
\begin{equation}
    \bar{M}=\left<\frac{j (j+1) \left(j^2+j-8\right)}{\mu ^5}\right>
\end{equation}
is wrong because the singular term $-2 B/(\gamma+1)$ can emerge only as a resummation of infinite number of powers of $q_0$ in the right-hand side of the dispersion relations\footnote{In the case of stronger singularities, such as graviton pole, one cannot take $\gamma\rightarrow -1$ because the integration with such a smearing function would lead to a divergence. This way, the allowed space of smearing functions encodes an information about the IR singularities in EFT.}. However, the situation is different if $B=0$. In this case 
\begin{equation}
    \bar{M}=\bar{M}^{(0)}+(\gamma+1)\bar{M}^{(1)},\quad \left<\bar{F}_{n4}\right>=\left<\bar{F}_{n4}^{(0)}\right>+(\gamma+1)\left<\bar{F}_{n4}^{(1)}\right>.
\end{equation}
For $B=0$ $\bar{M}^{(0)}$ doesn't depend on $\gamma$, as well as $\bar{F}_{n4}^{(0)}$. Thus, one can also separate the $\gamma$-dependent terms $\bar{M}^{(1)}$ and $\left<\bar{F}_{n4}^{(1)}\right>$ for example, by taking the derivative with respect to $\gamma$. The series in the right-hand side must correspond to a finite $\bar{M}^{(1)}$. Recall that we know about the finitness of it from an EFT computation, so it is a requirement of the IR theory imposed on the UV theory. Namely, the term $\left<\bar{F}_{n4}^{(1)}\right>$ must be also finite both for $\gamma\rightarrow -1$ and $q_0\rightarrow 0$ limits. This implies the finitness of the expression
\begin{equation}
    \left<\frac{j (j+1) \left(j^2+j-8\right)}{\mu ^5}\right>, 
\end{equation}
and enforces its relation to $\bar{M}^{(0)}$,
\begin{equation}
    \bar{M}^{(0)}=\left<\frac{j (j+1) \left(j^2+j-8\right)}{\mu ^5}\right>.
\end{equation}
Recall that we used the conditions that both $\bar{M}^{(0)}$ and $\bar{M}^{(1)}$ are finite for $\gamma\rightarrow -1$ (while $\gamma+1>0$). Actually, finitness of this limit is an implication of the finite forward limit $t\rightarrow 0$ for the original EFT expression. Let us note here that the non-analytic terms which still have smooth forward limit, such as $t \log{-t}$ or $\sqrt{-t}$ would not affect the relation between arcs and moments.

To conclude, if the dispersion relations contain non-analytic singular terms in the forward limit, there is no good relation between the EFT arc integrals and moments. However, if one can construct combinations of arc integrals which are finite in the forward limit, these combinations can be related to moments. The required conditions for holding this link are that after smearing of the arcs there is a smooth limit $\gamma\rightarrow -1$, and the arcs are differentiable in $\gamma$. Recall that it is important to keep small but finite $q_0$, in order to make all dispersion relations well-defined in a distributional sense. At the end one can notice that $q_0$ dependence would drop out after taking the limit $\gamma\rightarrow -1$.

The statement obtained here can be used in a non-trivial way for obtaining the EFT constraints in the presence of the forward limit singularities. Namely, the result tells that if one can construct any IR-finite combination (in this work we used linear combinations of $n_4$, $n_5$ and $n_6$ null-constraints weighted with the corresponding EFT couplings) this combination can be mapped to the corresponding combination of moments. This way, certain combinations of moments become entangled by null-constraints which can lead to compact bounds even if the original null-constraints are lost and cannot be taken to be zero or written in a form of moments. 

\section{Moment problem conditions \label{sec:C}}
In this appendix, we will give a brief introduction of inequalities we use in the analytic method in Sec.~\ref{sec5.1}. We refer the reader to~\cite{Chiang:2021ziz} for a systematical way of deriving these inequalities.

First, define the moment using the bracket
\begin{equation}
    a_{k,q}=\left<\frac{\mathcal{J}^{2q}}{\mu^k}\right>.
\end{equation}
The bracket means a integration over $\mu$ and summation over $\mathcal{J}$ with positive measure. To obtain compact bounds on these moments we use three types of inequalities.

1. {\it Relaxing inequalities.} They originate from the fact that the integrals in moments definition \eqref{brackets} start from $\epsilon^2$, setting $\epsilon^2=1$, 
\begin{equation}
\left<\frac{1}{\mu^n}\right> > \left<\frac{1}{\mu^{n+1}}\right>.
\end{equation}
Hence, we have
\begin{equation}
    a_{2,0}- a_{3,0}\geq0\,,\quad a_{3,0}- a_{4,0}\geq 0\,,\quad a_{3,1}- a_{4,1}\geq0\,,\dots  \,.\label{form:cons1}
\end{equation}

2. {\it Gram's inequalities.}
Calling $f_1(\mu,\mathcal{J}^2)=1/\mu^{3/2},\, f_2(\mu,\mathcal{J}^2)=1/\mu^{5/2}, \, f_3(\mu,\mathcal{J}^2)=\mathcal{J}^2/\mu^{5/2},$ we can obtain from the Gram's inequality \cite{20141071},
\begin{equation}
     \left|\begin{matrix}
        \int f_1^2 & \int f_1f_2& \int f_1f_3  \\[1.5mm]
        \int f_2f_1 & \int f_2^2& \int f_2f_3 \\[1.5mm]
        \int f_3f_1 & \int f_3f_2& \int f_3^2 
    \end{matrix}\right|\geq0\quad\Rightarrow\quad
    \left|\begin{matrix}
        a_{2,0} & a_{3,0}& a_{3,1}  \\[1.5mm]
        a_{3,0} & a_{4,0}& a_{4,1} \\[1.5mm]
        a_{3,1} & a_{4,1}& a_{4,2} 
    \end{matrix}\right|\geq0   \,,  \label{form:cons2}
\end{equation}
where we omitted the measure ${\rm d}\mu$ and the integration range $[\epsilon^2,\infty]$ in the above integrals.
In addition, we also have the rank-2 version of Gram's inequalities,
\begin{equation}
    \left|\begin{matrix}
        a_{2,0} & a_{3,0}  \\[1.5mm]
        a_{3,0} & a_{4,0} \\
    \end{matrix}\right|\geq0\,,\qquad 
    \left|\begin{matrix}
        a_{2,0} & a_{3,1}  \\[1.5mm]
        a_{3,1} & a_{4,2} 
    \end{matrix}\right|\geq0\,,\qquad
    \left|\begin{matrix}
        a_{4,0} & a_{4,1}  \\[1.5mm]
        a_{4,1} & a_{4,2} 
    \end{matrix}\right| 
    \geq0\,.   \label{form:cons3}
\end{equation}

 3. {\it Polytope boundary condition.} This condition follows from the fact that $\mathcal{J}^2=j(j+d-3)$ where $j\in\{0,2,4,6...\}$, is discrete-valued, therefore we have $(\mathcal{J}^2-6)(\mathcal{J}^2-20)\geq0$ in 4D for all possible values of $\mathcal{J}^2$ in the discrete set. Thus, we obtain
\begin{equation}
  \left<\frac{(\mathcal{J}^2-6)(\mathcal{J}^2-20)}{\mu^{5}}\right>\geq 0\qquad\Rightarrow \quad a_{4,2}-26a_{4,1}+120a_{4,0}\geq0.\label{form:cons4}
\end{equation}

These bounds along with the null constraints coming from the full crossing symmetries~\cite{Tolley:2020gtv,Caron-Huot:2020cmc}(see Eq.~\eqref{tree n4} for tree level) can already give a quiet good bound. But in order to obtain a better bound, we have to go to $k=5$ order. In principle, $k=5$ bounds can also come from the three types of inequalities mentioned above. We will list it here:
\begin{equation}
\begin{gathered}
\left(
\begin{array}{ccc}
a_{3,0} & a_{4,0} & a_{4,1} \\
a_{4,0} & a_{5,0} & a_{5,1} \\
a_{4,1} & a_{5,1} & a_{5,2}
\end{array}
\right),
\quad
\left(
\begin{array}{cccc}
a_{3,1} & a_{4,1} & a_{4,2} \\
a_{4,1} & a_{5,1} & a_{5,2} \\
a_{4,2} & a_{5,2} & a_{5,3}
\end{array}
\right)
\geq 0,\\
\left(
\begin{array}{ccccc}
a_{2,0} - a_{3,0} & a_{3,0} - a_{4,0} & a_{3,1} - a_{4,1} \\
a_{3,0} - a_{4,0} & a_{4,0} - a_{5,0} & a_{4,1} - a_{5,1} \\
a_{3,1} - a_{4,1} & a_{4,1} - a_{5,1} & a_{4,2} - a_{5,2}
\end{array}
\right),
\quad
\left(a_{4,0} - a_{5,0}\right),
\quad
\left(a_{4,1} - a_{5,1}\right)
\geq 0,\\
a_{5,2}-(6d+2)a_{5,1}+8(d^2-1)a_{5,0}\geq0 \qquad a_{5,3}-(6d+2)a_{5,2}+8(d^2-1)a_{5,1}\geq0
\end{gathered}
\end{equation}

These are bounds on a-geometry, we need to map this a-geometry to g-geometry, this is where the non-linearity comes in (see Eq.~(\ref{4dnull}--\ref{d6g4})). The non-linearity makes projecting out irrelevant moments or Wilson coefficients analytically difficult to proceed. The key observation is that k=5 conditions modifies only the "kink" part of $\tilde{g}_3-\tilde{g}_4$ bounds~\cite{Chiang:2021ziz}. So the effect of $k=5$ inequalities correspond to a set of equalities characterizing boundaries in the "kink" part.

\begin{equation}
\begin{gathered}
\rm{Rank}
\left(
\begin{array}{ccc}
a_{3,0} & a_{4,0} & a_{4,1} \\
a_{4,0} & a_{5,0} & a_{5,1} \\
a_{4,1} & a_{5,1} & a_{5,2}
\end{array}
\right)
= 2, \quad
\rm{Rank}
\left(
\begin{array}{ccc}
a_{3,1} & a_{4,1} & a_{4,2} \\
a_{4,1} & a_{5,1} & a_{5,2} \\
a_{4,2} & a_{5,2} & a_{5,3}
\end{array}
\right)
= 2,\\
\rm{Rank}
\left(
\begin{array}{ccc}
a_{2,0} - a_{3,0} & a_{3,0} - a_{4,0} & a_{3,1} - a_{4,1} \\
a_{3,0} - a_{4,0} & a_{4,0} - a_{5,0} & a_{4,1} - a_{5,1} \\
a_{3,1} - a_{4,1} & a_{4,1} - a_{5,1} & a_{4,2} - a_{5,2}
\end{array}
\right)
= 1,\\
a_{4,2}-(6d+2)a_{4,1}+8(d^2-1)a_{4,0}=0,
\\a_{5,2}-(6d+2)a_{5,1}+8(d^2-1)a_{5,0}=0, \\ a_{5,3}-(6d+2)a_{5,2}+8(d^2-1)a_{5,1}=0.
\end{gathered}
\end{equation}

%

\bibliographystyle{utphys}
\bibliography{ref}


\end{document}